\documentclass[prl,twocolumn,superscriptaddress]{revtex4-2}
\usepackage[utf8]{inputenc}
\usepackage[american]{babel}
\usepackage[pdftex]{graphicx}  
\usepackage{graphicx, xcolor}
\usepackage{dcolumn}
\usepackage{bm}
\usepackage{amsmath,amsthm,amssymb}
\usepackage{hyperref}
\usepackage{xcolor}
\hypersetup{colorlinks,bookmarksopen,bookmarksnumbered,
citecolor=red,
linkcolor=red,
pdfstartview=false,
urlcolor=red}
\usepackage{graphicx}

\usepackage{braket}

\usepackage[normalem]{ulem}

\begin{document}
\newcommand{\titleinfo}{Error-resilience Phase Transitions in Encoding-Decoding Quantum Circuits}

\title{\titleinfo}

\author{Xhek Turkeshi}
\affiliation{Institut f\"ur Theoretische Physik, Universit\"at zu K\"oln, Z\"ulpicher Strasse 77, 50937 K\"oln, Germany}
\affiliation{JEIP, USR 3573 CNRS, Coll\`ege de France, PSL Research University,
11 Place Marcelin Berthelot, 75321 Paris Cedex 05, France}
\author{Piotr Sierant}
\affiliation{ICFO-Institut de Ci\`encies Fot\`oniques, The Barcelona Institute of Science and Technology, Av. Carl Friedrich Gauss 3, 08860 Castelldefels (Barcelona), Spain}

\date{\today}

\begin{abstract}
Understanding how errors deteriorate the information encoded in a many-body quantum system is a fundamental problem with practical implications for quantum technologies. Here, we investigate a class of encoding-decoding random circuits subject to local coherent and incoherent errors. 
We analytically demonstrate the existence of a phase transition from an error-protecting phase to an error-vulnerable phase occurring when the error strength is increased.
This transition is accompanied by R\'enyi entropy transitions and by onset of multifractal features in the system. Our results provide a new perspective on storing and processing quantum information, while the introduced framework enables an analytic understanding of a dynamical critical phenomenon in a many-body system.
\end{abstract}
\maketitle





\paragraph{Introduction.} 
Comprehending the robustness of quantum many-body systems to errors is critical for near-term quantum computation~\cite{Preskill2018quantumcomputingin,fraxanet2022coming}. 
Indeed, despite the outstanding experimental advances~\cite{Arute2019,Zhong2020,wu2021strong,Huang2022,Madsen2022}, an unquestionable quantum advantage is likely to be achieved only with scalable and fault-tolerant computers~\cite{nielsen00,gottesman1997stabilizer}. 
Besides, error-correcting features provide an insightful perspective on the many-body problem, 
leading to phenomena that emerge due to the interplay of errors and the intrinsic dynamics of the system.
Notable examples include measurement-induced phase transitions in monitored quantum matter~\cite{Fisher2023,potter2022quantumsciencesandtechnology,lunt2022quantumsciencesandtechnology,skinner2019measurementinducedphase,li2019measurementdrivenentanglement,jian2020measurementinducedcriticality,gullans2020dynamicalpurificationphase,bao2020theoryofthe,zabalo202criticalpropertiesof,sierant2022measurementinducedphase,noel2021measurementinducedquantum,koh2022experimentalrealizationof,hoke2023quantuminformationphases,behrends2022surface,Venn2023coherent}, complexity phase transitions in the sampling of random circuits~\cite{morvan2023phase,ghosh2023complexity,ware2023sharp}, or non-stabilizer phase transitions~\cite{leone2023phase,niroula2023phase}.
In these scenarios, the system's resilience to errors is reflected in non-local structural properties, such as entanglement or other non-linear functions of the density matrix~\cite{fisher2023randomquantumcircuits}. Consequently, analytical results are sparse, and identifying solvable many-body setups highlighting error-correcting features is an outstanding problem. 

\begin{figure}[b!]
    \centering
    \includegraphics[width=\columnwidth]{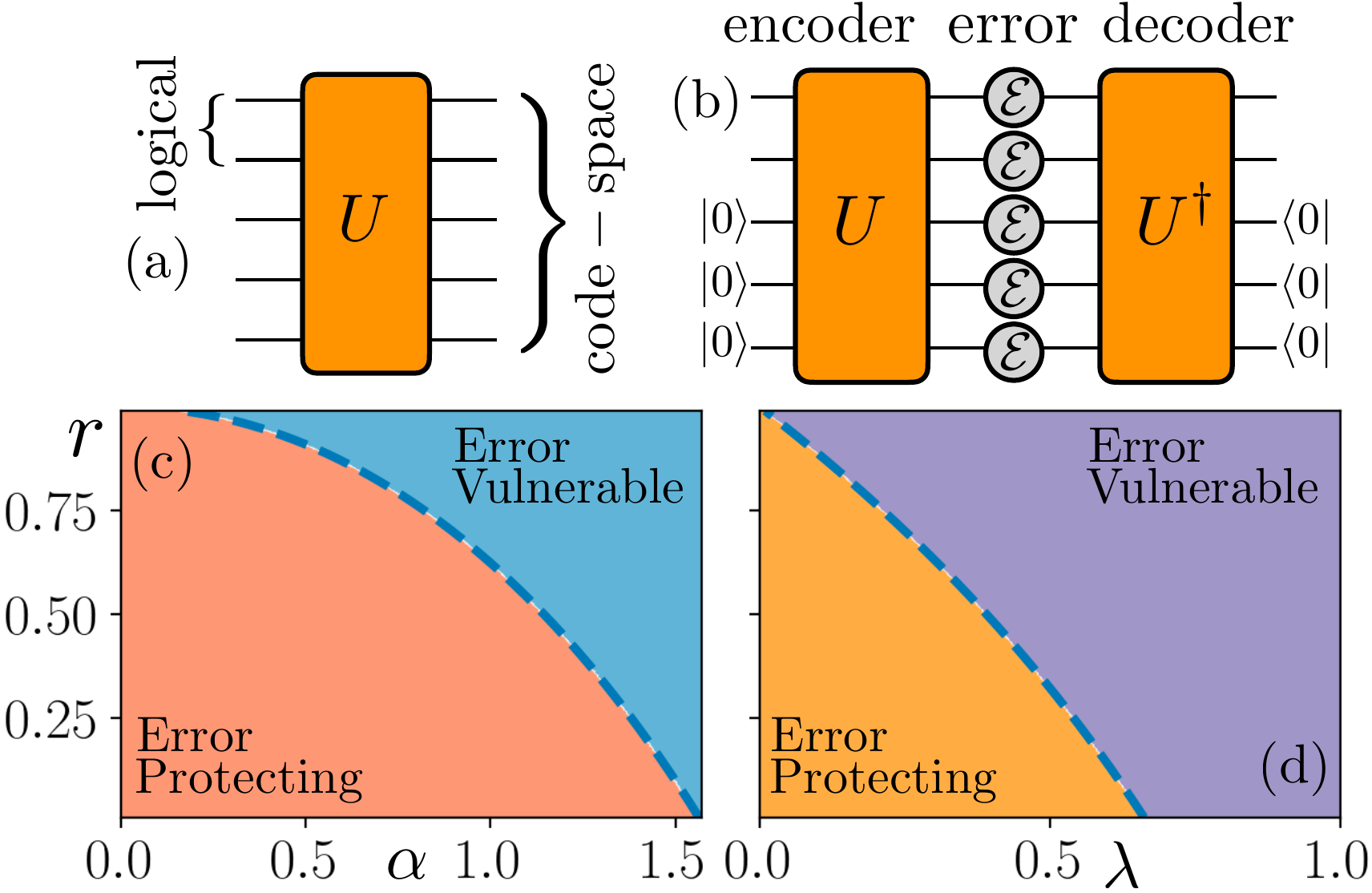}
    \caption{(a) The unitary $U$ encodes the logical state of $k$ qubits onto the code space with $N$-qubits. (b) The circuit architecture comprises encoder $U$, a layer of local errors $\mathcal{E}_i$, and decoder  $U^\dagger$. The decoded logical state is obtained from the code-space by projecting the ancilla qubits onto ${|0_{\bar{X}}\rangle}$. 
    (c) Phase diagram over the code rate $r=k/N$ and the error strength $\alpha$ for the coherent errors. (d) Phase diagram for the depolarizing channel of strength $\lambda$. The dashed blue lines denote the critical line separating the error-protecting and error-vulnerable phases.
    }
    \label{fig:cartoon}
\end{figure}

This work presents an exactly solvable class of random quantum circuits composed of an encoding unitary $U$, a layer of local errors described by a quantum channel  $\mathcal{E}$, and a decoding unitary $U^\dagger$, cf. Fig.~\ref{fig:cartoon}. We separately consider: i) coherent errors, in the form of rotations by an angle $\alpha$~\cite{Bravyi_2018}, ii) incoherent errors modeled by local depolarizing channels of strength $\lambda$~\cite{nielsen00}. 
We demonstrate that the decoded logical state of this circuit undergoes an \textit{error-resilience phase transition} between an error-protecting phase (\textbf{EPP}) and an error-vulnerable phase (\textbf{EVP}). In the EPP, The decoded logical state  is asymptotically close to the initial state. On the contrary, in the EVP, the logical state becomes mixed for incoherent errors or can be retrieved only with a complex global unitary operation for coherent errors.
We identify the critical point ${\alpha=\alpha_c}$ (${\lambda=\lambda_c}$) of the transition for coherent (incoherent) error channel by quantifying the fidelity between the decoded and initial logical states and characterize the EPP and EVP by the decoded logical state structure. 
Our results are based on exact analytical calculations valid for all values of parameters at any system size and are benchmarked with numerical computations.

\paragraph{Encoding-Decoding Circuits.} 
The system of interest is composed of $k$ logical qubits (a subsystem $X$)  and additional ${N-k}$ ancilla qubits (denoted by $\bar{X}$), which, together, form a $2^N$-dimensional code-space.
The initial state of the logical qubits is $|\psi_X\rangle$ and, without loss of generality, the ancilla qubits are initialized in the state ${\ket{0_{\bar X}} \equiv|0\rangle^{\otimes (N-k)}}$. The encoding unitary $U$ is chosen with the Haar measure from the unitary group $\mathcal U(2^N)$. 
Its action is followed by the layer of local errors ${\mathcal{E}=\mathcal{E}_1\circ \dots \circ \mathcal{E}_N}$ with $\mathcal{E}_i$ acting on the $i$-th site, and the decoding unitary $U^\dagger$. 
The coherent error channel is specified by ${\mathcal{E}_i(\rho) = K_{0,i} \rho K_{0,i}^\dagger}$ with ${K_{0,i}= e^{-i \alpha \sigma_i^z/2}}$ a local rotation by an angle ${\alpha\in [0,\pi/2]}$~\footnote{We denote $\sigma^{x,y,z}_i$ the Pauli matrices acting on the $i$-th qubit.}. 
Separately, we consider the incoherent channel 
with local depolarizing noise, ${\mathcal{E}_i(\rho) = \sum_{\mu=0}^3 K_{\mu,i}\rho K^{\dag}_{\mu,i}}$, where Kraus operators read $K_{0,i} = \sqrt{1-3\lambda/4}$, $K_{1,i} = \sqrt{\lambda/4} \sigma^x_i$, $K_{2,i} = \sqrt{\lambda/4} \sigma^y_i$, $K_{3,i} = \sqrt{\lambda/4} \sigma^z_i$ and $\lambda \in [0,1]$ fixes the error strength. 
The global error channel $\mathcal{E}$ is specified
as $\mathcal E(\rho) = \sum_{\vec{\mu}} K_{\vec{\mu}} \rho K^{\dag}_{\vec{\mu}}$
with the Kraus operators $K_{\vec{\mu}}= \prod_{j=1}^N K_{\mu_j,j}$, 
with $\vec{\mu}=(\mu_1,\ldots, \mu_N)$ ranging over $4^N$ values for the incoherent errors and $\vec{\mu}=(0,\ldots, 0)$ in the coherent case.
Denoting the initial state as ${\rho_0=|(\psi_X, 0_{\bar{X}})\rangle\langle  (\psi_X, 0_{\bar{X}})|}$, the resulting state in the code-space is ${\rho = U^\dagger \mathcal{E}(U \rho_0 U^\dagger )U}$. 
The final decoded logical state is obtained by projecting out the ancilla qubits ${\rho_X = \langle 0_{\bar{X}}|\rho|0_{\bar{X}}\rangle/\mathrm{tr}(\langle 0_{\bar{X}}|\rho|0_{\bar{X}}\rangle)}$. 

When no error is present (${\alpha=0}$ or ${\lambda=0}$), we retrieve the initial logical state, ${\rho_X= |\psi_X\rangle\langle \psi_X|}$. The retrieval becomes more difficult with increasing error strength. In the following, we show that the competition between the errors and the decoding action of the circuit leads to a transition between the EPP, in which the logical state is recovered, and the EVP, in which the information about $|\psi_X\rangle$ gets scrambled over the logical subspace.
To this end, we consider the fidelity ${F=\langle \psi_X |\rho_X|\psi_X\rangle}$ between the initial (pure) and the final logical states~\cite{nielsen00}. We also study the R\'enyi entropy with index $q$ of the decoded logical state $\rho_X$, namely 
\begin{equation}
\begin{split}
    S_q &= \frac{1}{1-q}\log_2[\mathrm{tr}_{X_1} \left(\mathrm{tr}_{{X_2}}
    [\rho_X]\right)^q]
\end{split}\label{eq:defobs1}
\end{equation}
where ${X=X_1\cup X_2}$. For the coherent errors, \eqref{eq:defobs1} corresponds to the entanglement entropy for a bipartition of $X$ into subsystems $X_1$ and $X_2$. For the incoherent errors, we set $X_2$$=$$\emptyset$ and $X_1$$=$$X$ to study whether $\rho_X$ is mixed.

\paragraph{The replica trick.}
All the above quantities for both types of errors can be unified within a single framework based on the following replica trick. We consider 
\begin{equation}
    \begin{split}
    {A}_{U,\mathcal{E}}^{(2q)} &= (U^\dagger)^{\otimes2q} \mathcal{K}^{(2q)} U^{\otimes 2q}\\
    \mathcal{K}^{(2q)} & \equiv  \sum_{\vec{\mu}^{(1)},\dots,\vec{\mu}^{(q)}}\bigotimes_{m=1}^q \left(K_{\vec{\mu}^{(m)}}\otimes K^\dagger_{\vec{\mu}^{(m)}}\right),
    \label{eq:zioau}
    \end{split}
\end{equation}
which acts on $2q$ copies of the system, with each $K_{\vec{\mu}^{(m)}}$ and $K^\dagger_{\vec{\mu}^{(m)}}$ acting on a separate replica. 
Non-linear functions $\Lambda$ of the density matrix can be calculated as functionals $\Phi^{(2q)}_\Lambda ({A}_{U,\mathcal{E}}^{(2 q)})$ with appropriately chosen number of replicas $q$. 
We detail derivations for all considered quantities 
in~\cite{supmat}. 
As an example, by putting $q=1$, we can rewrite the fidelity $F$ between $\ket{\psi_X}\bra{\psi_X}$ and $\rho_X$ as 
\begin{equation}
 \Phi_{F}^{(2)}
 =  \frac{\langle (\psi_X,0_{\bar{X}}),(\psi_X,0_{\bar{X}})|{A}_{U,\mathcal{E}}^{(2)}| (\psi_X,0_{\bar{X}}),(\psi_X,0_{\bar{X}})\rangle}{\sum_x\langle (\psi_X,0_{\bar{X}}),(x,0_{\bar{X}})|{A}_{U,\mathcal{E}}^{(2)}| (x,0_{\bar{X}}),(\psi_X,0_{\bar{X}})\rangle},
 \label{eq:example}
\end{equation}
where the sum is over $2^k$ basis states $\ket{x}$ of the logical subspace. Since $U$ is randomly selected from $\mathcal U(2^N)$, we consider both quenched,  $\overline \Lambda
\equiv \mathbb{E}_{U}[\Phi^{(2q)}_\Lambda({A}_{U,\mathcal{E}}^{(2 q)})]$, and annealed, $
\tilde \Lambda
\equiv \Phi^{(2q)}_\Lambda[\mathbb{E}_{U}({A}_{U,\mathcal{E}}^{(2 q)})]$, averages $\mathbb{E}_{U}(.)$ over the realizations of the circuit. 
The Schur-Weyl duality reads
\begin{equation}
    \mathbb{E}_{U}\left({A}_{U,\mathcal{E}}^{(2 q)} \right) \equiv  \mathcal{A}_\mathcal{E}^{(2q)} 
    = \sum_{\pi \in \mathcal{S}_{2q}} b_\pi(\mathcal{E}) T_\pi,
    \label{eq:schur-weyl}
\end{equation}
where $\mathcal{S}_{2q}$ is the permutation group of $2q$ elements, $T_\pi$ is a permutation representation in the replica space, and $b_\pi(\mathcal{E}) =  \sum_\sigma W_{\pi,\sigma} \mathrm{tr}( \mathcal{K}^{(2q)} T_\sigma)$ is fixed by the error model $\mathcal{E}$ and the Weingarten symbols $W_{\pi,\sigma}$~\cite{Roberts_2017}. 
The calculation of the annealed averages 
$\tilde \Lambda$ with \eqref{eq:schur-weyl} requires computation of objects of the form $I_{\mathcal{B}}= \mathrm{tr}(\mathcal{A}^{(2q)}_{\mathcal{E}}\mathcal{B}({2q}))$ with certain boundary conditions $\mathcal{B}({2q})$. 
For example, numerator and denominator in the second term of~\eqref{eq:example} are respectively fixed by the replica boundary conditions ${\mathcal{B}_\mathrm{num}^F\equiv \rho_0^{\otimes 2}}$ and  $\mathcal{B}_\mathrm{den}^F=\rho_0\otimes (\openone_X \otimes |0_{\bar{X}}\rangle\langle 0_{\bar{X}}|)T_{(1,2)}$, with the permutation $(1,2)$ swapping the two replicas.

\paragraph{Numerical implementation.}
The outlined reasoning allows us to obtain exact formulas for the annealed averages of the fidelity, $\tilde F$, and of the entropies $\tilde S_q$, which we test by performing numerical simulations of the encoding-decoding circuits.
The encoding unitary $U$ comprises $T$ layers of local 2-qubit gates randomly selected from $\mathcal{U}(4)$ and arranged in a brick-wall pattern with periodic boundary conditions.
The initial logical state is set to ${|\psi_X\rangle = |0_X\rangle\equiv |0\rangle^{\otimes k}}$. 
The considered circuit is unitary for coherent errors, implying that the final logical state $\rho_X$ is \emph{pure}, allowing simulations for up to $N=24$ qubits. 
In contrast, incoherent errors result in a mixed logical state, limiting our numerics to $N\le 12$. 
We calculate the quenched averages $\overline{F}$, $\overline{S_q}$ by averaging the results over more than $1000$ realizations of the circuit. Throughout, we fix the circuit depth $T=2N$ and note that for large $N$ other choices $T\propto N$ yield the same results,~cf.~\cite{brandao2016local, supmat}.
Lastly, we anticipate that the quantities of interest are self-averaging, and the discrepancies between the annealed $\tilde{\Lambda}$ and quenched averages $\overline{\Lambda}$ are exponentially suppressed in $N$. Hence, our exact solutions for the annealed averages fully capture the physics of encoding-decoding circuits in the scaling limit $N\to \infty$.

\begin{figure}
    \centering
    \includegraphics[width=\columnwidth]{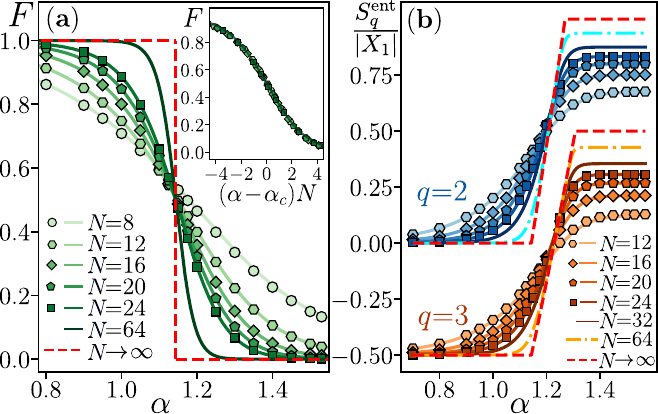}
    \caption{Error-resilience transition for coherent errors at code rate $r=1/2$. (a) Fidelity $F$ for varying error strength $\alpha$. The markers show the the quenched averages $\overline{F}$, while the solid lines represent the exact result~\eqref{eq:exactdist} for the annealed average $\tilde{F}$.  
    (Inset) The critical behavior around $\alpha=\alpha_c$ with the critical exponent $\nu=1$. (b) Entanglement entropy of the decoded logical state,
    for a subsystem of size $|X_1| = N/4$ for $q=2$ and $q=3$ (shifted by a constant for clarity). Markers and solid lines correspond to the annealed averages (numerical data) and the quenched averages (analytic expressions, see~\cite{supmat}).
    The error bars are smaller than the marker size.
     }
    \label{fig:fig2ALT}
\end{figure}

\paragraph{Error-resilience phase transitions for coherent errors.}
We start by considering the coherent error case. Performing the annealed average of~\eqref{eq:example} over $\mathcal U(2^N)$, we find  
\begin{equation}
    \tilde{F} = \frac{\left(2^{N} - 1\right) \left(2^{N} \cos^{2N}(\alpha/2)+ 1\right)}{2^{N} \cos^{2N}(\alpha/2)\left( 2^{N} - 2^{k}\right) + 2^{N + k} - 1}.\label{eq:exactdist}
\end{equation}
In Fig.~\ref{fig:fig2ALT}(a), we compare this analytical formula for $\tilde{F}$ with the numerically computed quenched average $\overline{F}$. 
A quantitative agreement between $\tilde{F}$ and $\overline{F}$ is observed already for ${N=8}$, and the quenched and annealed fidelitys approach each other exponentially fast in $N$~\cite{supmat}. 
A phase transition between EPP and EVP is revealed upon extrapolation of \eqref{eq:exactdist} to the thermodynamic limit, which demonstrates that $\tilde{F}$ develops a step function behavior typical for order parameters 
\begin{equation}
    \lim_{k,N \to \infty,r=\mathrm{const}} \tilde{F} = \Theta\left(\alpha < \alpha_c(r)\right),\label{eq:thetafun}
\end{equation}
where $r=k/N$ is the code rate, $\alpha_c(r)=2 \arccos(2^{-r/2})$ is the critical point, and $\Theta$ is the Heaviside step function. Moreover, we observe the emergence of critical behavior $\tilde{F}=f( (\alpha-\alpha_c)N^{1/\nu})$, where $f$ is a universal function and $\nu=1$ is the critical exponent, cf. Fig.~\ref{fig:fig2ALT}(a).
Finally, from~\eqref{eq:thetafun}, we obtain the phase diagram summarized in Fig.~\ref{fig:cartoon}(c). 
Below the critical strength $\alpha_c(r)$, the logical subspace is protected from the errors, and the initial state $|\psi_X\rangle$ is faithfully recovered. In contrast, for $\alpha>\alpha_c(r)$, the decoded state $\rho_X$ is orthogonal to $\ket{\psi_X}\bra{\psi_X}$. 
Again, being unitary, the
error can be corrected via a global unitary operation on the logical qubits. 
The above discussion demonstrates the presence of the error-resilience transition at critical error strength $\alpha_c$ that separates the EPP and EVP. 
In the following, we further characterize the properties of the two phases for coherent errors.

\paragraph{Error-protecting and error-vulnerable phases.}
To characterize the properties of the EPP and EVP, we investigate the entanglement entropy~\eqref{eq:defobs1} of the decoded logical (pure) state $\rho_X$ varying ${\alpha\in[0,\pi/2]}$. Concretely, we consider the initial logical state $\ket{\psi_X} = |0\rangle^{\otimes k}$, but what follows applies to more general initial conditions~\cite{supmat}.
As for the fidelity, the entanglement entropy $S_q$ can be written as a replica functional over $2q$ copies of the system. We detail the computations and the exact, but lengthy, closed expressions in~\cite{supmat,turkeshi_2023_10302870}. 
In the scaling limit, we find that $\tilde{S}_q = a_q |X_1|$, with 
\begin{equation}
    a_q = \begin{cases} 0,& \alpha<\alpha_c(r)\\
    \frac{2 q \left(\log _2\left[\cos \left(\alpha/2\right)\right]+(1-r)/2\right)}{(1-q) r_1}, & \alpha_c(r)<\alpha<\alpha^\mathrm{ent}_X(r),\\
    1,&\text{otherwise},
    \end{cases} \label{eq:entvolX}
\end{equation}
where 
$r_1\equiv |X_1|/N\le r/2$ and $|X_1|$ is the size of the smallest subsystem in the bipartition of $X$.
This exact expression implies that the transition between EPP and EVP manifests as an area (${a_q=0}$) to volume-law (${a_q>0}$) entanglement phase transition of the logical state, similar, in certain aspects, to the transition observed in monitored quantum circuits \cite{li2019measurementdrivenentanglement,skinner2019measurementinducedphase}. 
Notably, the volume-law scaling of $\tilde{S}_q^\mathrm{ent}$ for $\alpha>\alpha_c(r)$ indicates that a global unitary operation on the logical qubits is required to retrieve the initial state $\ket{\psi_X}$. 
The error strength for which $\tilde{S}_q^\mathrm{ent}$ scales with the prefactor $a_q=1$ equal to unity depends non-trivially on the R\'enyi index $q$, $\alpha_X^\mathrm{ent}(r_1)=2 \arccos(2^{\frac{q (r-r_1-1)+r_1}{2 q}})$. Hence, our results demonstrate the existence of an extended interval of error strength within the EVP, $\alpha_c(r)<\alpha<\alpha^\mathrm{ent}_X(r)$, in which the logical state is \emph{multifractal}~\cite{stephan2009,luitz2014universalbehaviorbeyond,lindinger2019manybodymultifractality,mace2019multifractal,Pietracaprina2021, roy2022hilbertspace,
sierant2022universalbehaviorbeyond}. 
The annealed average of the entanglement entropy $\tilde{S}_q$ is in a quantitative agreement with the quenched averages $\overline{S}_q$, as shown in Fig.~\ref{fig:fig2ALT}(b). Moreover, the entanglement entropy follows the critical scaling of the fidelity, i.e., in the vicinity of the phase transition $S_q=g( (\alpha-\alpha_c)N^{1/\nu})$ where $\nu=1$ is the critical exponent and $g$ is a universal function.
These considerations extend to further observables, including the participation entropies which reveal unique localization features 
both for the logical and code-space states~\cite{supmat}.
We 
conclude this section briefly summarizing the self-averaging properties of the considered quantitites across the error-resilience transition. 
In \cite{supmat}, we demonstrate that the differences between the annealed and quenched averages of the fidelity $F$ and the entanglement entropy \eqref{eq:defobs1} decay exponentially with the number of qubits $N$. Each considered quantity involves a functional over the replica space with a numerator and denominator. 
There, we argue that the self-averaging is caused by an exponential decrease $\sim \exp(-\gamma N)$ of the relative realization-to-realization fluctuations of the numerators and denominators, where $\gamma$ depends on the observable and the parameters of the circuit.

\begin{figure}
    \centering
    \includegraphics[width=\columnwidth]{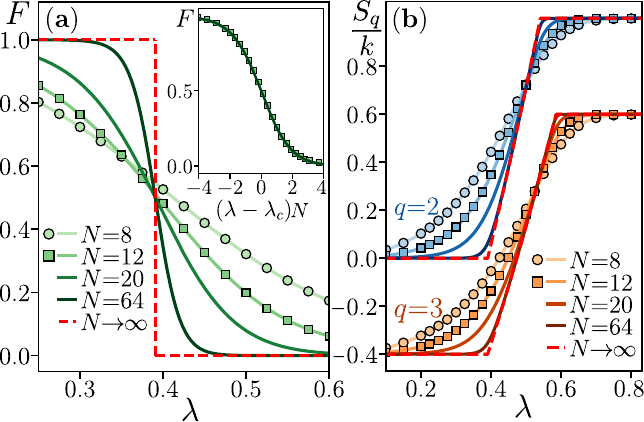}
    \caption{Error-resilience phase transition for depolarizing noise at $r=1/2$. 
    (a) Fidelity $F$ between initial and final logical state $\rho_X$ for varying error strength $\lambda$; the markers denote quenched averages, the solid lines represent the exact results~\eqref{eq:exlambdF}.
    (Inset) Data collapse for $\nu=1$ around $\lambda=\lambda_c$. (b) (Thermodynamic) entropy of $\rho_X$ for $q=2$ and $q=3$ (shifted by a constant); the markers denote the quenched averages, the solid lines are the annealed averages~\cite{supmat}. Error bars are smaller than the marker size.
    }
    \label{fig:fig3}
\end{figure}

\paragraph{Error-resilience phase transitions for incoherent errors.} 
The 
phenomenology of error-resilience phase transition extends to other noise models, including incoherent error sources. We exemplify this fact by considering the layer of depolarizing noise, 
as previously defined. The considerations and computations parallel the coherent case, 
see~\cite{supmat}. 
There are two compelling differences: (i) the evaluation of the coefficients $b_\pi(\mathcal{E})$ is more involved, and (ii) the decoded logical state $\rho_X$ is, in general, mixed.
We obtain the exact expression for the annealed average of the fidelity, ${F=\langle \psi_X |\rho_X|\psi_X\rangle}$, namely
\begin{equation}
  \tilde{F} = \frac{\left(2^{N} - 1\right) \left(2^N \left(1 - 3 \lambda/4 \right)^{N}+1\right)}{ 2^N \left(1 - 3 \lambda/4\right)^{N} \left(2^{N} - 2^{k}\right)  + 2^{ N + k} - 1  }\label{eq:exlambdF}.
\end{equation}
As we show in Fig.~\ref{fig:fig3}(a) the annealed average quantitatively agrees with the numerical calculated quenched average $\overline{F}$ already for ${N=8}$. Interestingly, the discrepancy between $\tilde{F}$ and $\overline{F}$ is significantly smaller than in the coherent case, suggesting the self-averaging is even stronger for the incoherent errors~\cite{supmat}.
Taking the scaling limit, we retrieve the $\Theta$-function behavior 
\begin{equation}
	\lim_{k,N\to\infty,r=\mathrm{const}}\tilde{F} = \Theta\left(\lambda < \lambda_c(r)\right),\label{eq:ziobono1}
\end{equation}
highlighting the order parameter nature of $\tilde{F}$ and the position of the critical point $\lambda_c(r) = 4\left(1-2^{r-1}\right)/3$. The resulting phase diagram is summarized in Fig.~\ref{fig:cartoon}(d). 
We note that even at zero code rate, $r=0$, an error-resilience transition is present, in contrast to coherent errors that are always error-resilient at $r=0$, highlighting the fundamentally different nature of incoherent noise. In the EPP, for $\lambda < \lambda_c$, the logical state is faithfully recovered, hence $\rho_X$ remains pure in the 
thermodynamic limit. In contrast, at $\lambda>\lambda_c(r)$, 
$\rho_X$ becomes a mixed state.

To highlight the physical properties of the EVP, we study the thermodynamic entropy of $\rho_X$ by putting $X_2$$=$$\emptyset$ and $X_1$$=$$X$ in \eqref{eq:defobs1}. 
The technical details are reported in~\cite{supmat,turkeshi_2023_10302870}, 
while here we discuss only the scaling limit. We find $\tilde{S}_q = a_q k$ with
\begin{equation}
	a_q = \begin{cases} 0,& \lambda<\lambda_c(r)\\
   \frac{q \left(-\log _2(4-3 \lambda )+r+1\right)}{(q-1) r}, & \lambda_c(r)<\lambda<\lambda^\mathrm{th}_X(r),\\
    1,&\text{otherwise},
    \end{cases} \label{eq:thermo}
\end{equation}
where $\lambda^\mathrm{th}_X(r)=(4-2^{\frac{q+r}{q}})/3$ is the value at which $a_q$ reaches unity. 
The non-trivial dependence of $\lambda^\mathrm{th}_X(r)$ on $q$ highlights the multifractal nature of $\rho_X$ which becomes a maximally mixed state only in the limit $\lambda \to 1$.
For the incoherent errors, the error-resilience transition manifest itself as a purification transition~\cite{gullans2020dynamicalpurificationphase}, with $a_q=0$ for $\lambda <\lambda_c$ and $a_q>0$ for $\lambda>\lambda_c(r)$.

\paragraph{Conclusion and outlook.}
We have investigated a quantum circuit of an encoder-decoder architecture subject to a layer of 
local errors. 
The interplay of the errors with the intrinsic dynamics of the encoder-decoder circuit and the projection on ancilla qubits leads to a phase transition between EPP and EVP. Our exact analytic calculations allow us to pinpoint the phase transition and understand it in terms of the retrievability of the logical state. For coherent errors, we show that this transition manifests itself as a 
an area-to-volume law transition in the entanglement structure of the logical state, and is associated with a localization transition in the computational basis. 
In a parallel fashion, incoherent errors, despite being fundamentally different from the coherent rotations, induce an area-to-volume law transition in the thermodynamic entropy of the logical state. We also emphasize the multifractal features of the EVP.

Our results are of direct relevance for experiments on noisy quantum devices~\cite{niroula2023phase}. Indeed, in~\cite{supmat}, we numerically show the robustness of our findings when both encoder and decoder are implemented via noisy circuits. Moreover, in \cite{supmat}, we extend our methods to error models with site-dependent error strengths. We demonstrate the persistence of the phenomenology of the EPP-EVP transition and find that the disorder introduced to the system is a relevant perturbation in the renormalization group sense, changing the critical exponent to $\nu = 2$.

The renowned threshold theorems \cite{aharonov2008fault, Shor96FaultTolerant, knill1998resilient} establish threshold error strength below which the quantum information encoded in a larger code-space can be retrieved with certainty. While we find a similar phenomenology, our protocol differs by two essential aspects from the scenarios typically analyzed in the context of quantum error correction: (i) rather than finding an optimal decoder, we fix the unitary $U^\dag$ as the decoder; (ii) instead of measuring the ancillary qubits and implementing recovery operations, we consider a projection over the ancilla subsystem, in line with the idea of the post-selected quantum computing~\cite{Knill2005quantumcomputingwith}. These assumptions enable the analytical solution for the considered quantum information processing protocol which, due to the uncovered phase transition in dynamics of the system, is of interest also from the perspective of non-equilibrium many-body physics.

There are various possible extensions of this work. In a future contribution~\cite{sieranttoappear}, we report the non-stabilizerness properties of the considered circuits, generalizing the recent results~\cite{niroula2023phase} and extending our methods to Clifford circuits. Shallow circuits and errors interspersing with unitary gates appears as an important venue for future investigation. The similarity of the error-resilience transitions for the coherent and depolarizing errors suggests that our methods generalize to other noise models. A systematic analysis in this direction is desirable, particularly due to possible insights into the noise limitations of quantum computations~\cite{niroula2023error,fefferman2023effect,quek2023exponentially}. Finally, our methods can be extended to calculation of coherent quantum information~\cite{Schumacher96processing}, which provides insights into coding with quantum circuits~\cite{Gullans21Coding} without optimizing over decoders, cf.~\cite{colmenarez2023accurate, Bombin12resilience}.
We leave these considerations for further studies.

\begin{acknowledgments}
\paragraph{Acknowledgments.}
We thank R. Fazio, G. Fux, and S. Pappalardi for discussions. XT thanks A. Altland, L. Colmenarez, A. Rosch, and M. M\"uller for insightful discussions. 
X.T. acknowledges support from the ANR grant "NonEQuMat"
(ANR-19-CE47-0001) and DFG under Germany's Excellence Strategy – Cluster of Excellence Matter and Light for Quantum Computing (ML4Q) EXC 2004/1 – 390534769, and DFG Collaborative Research Center (CRC) 183 Project No. 277101999 - project B01. 
P.S. acknowledges support from: ERC AdG NOQIA; MICIN/AEI (PGC2018-0910.13039/501100011033, CEX2019-000910-S/10.13039/501100011033, Plan National FIDEUA PID2019-106901GB-I00, FPI; MICIIN with funding from European Union NextGenerationEU (PRTR-C17.I1): QUANTERA MAQS PCI2019-111828-2); MCIN/AEI/ 10.13039/501100011033 and by the "European Union NextGeneration EU/PRTR" QUANTERA DYNAMITE PCI2022-132919 within the QuantERA II Programme that has received funding from the European Union's Horizon 2020 research and innovation programme under Grant Agreement No 101017733Proyectos de I+D+I "Retos Colaboración" QUSPIN RTC2019-007196-7); Fundació Cellex; Fundació Mir-Puig; Generalitat de Catalunya (European Social Fund FEDER and CERCA program, AGAUR Grant No. 2021 SGR 01452, QuantumCAT \ U16-011424, co-funded by ERDF Operational Program of Catalonia 2014-2020); Barcelona Supercomputing Center MareNostrum (FI-2023-1-0013); EU (PASQuanS2.1, 101113690); EU Horizon 2020 FET-OPEN OPTOlogic (Grant No 899794); EU Horizon Europe Program (Grant Agreement 101080086 — NeQST), National Science Centre, Poland (Symfonia Grant No. 2016/20/W/ST4/00314); ICFO Internal "QuantumGaudi" project; European Union's Horizon 2020 research and innovation program under the Marie-Skłodowska-Curie grant agreement No 101029393 (STREDCH) and No 847648 ("La Caixa" Junior Leaders fellowships ID100010434: LCF/BQ/PI19/11690013, LCF/BQ/PI20/11760031, LCF/BQ/PR20/11770012, LCF/BQ/PR21/11840013). Views and opinions expressed are, however, those of the author(s) only and do not necessarily reflect those of the European Union, European Commission, European Climate, Infrastructure and Environment Executive Agency (CINEA), nor any other granting authority. Neither the European Union nor any granting authority can be held responsible for them.
\end{acknowledgments}

%

\bibliographystyle{apsrev4-2}

\widetext
\clearpage
\begin{center}
\textbf{\large \centering Supplemental Material:\\ Error-resilience Phase Transitions in Encoding-Decoding Quantum Circuits}
\end{center}

\setcounter{equation}{0}
\setcounter{figure}{0}
\setcounter{table}{0}
\setcounter{page}{1}
\renewcommand{\theequation}{S\arabic{equation}}
\setcounter{figure}{0}
\renewcommand{\thefigure}{S\arabic{figure}}
\renewcommand{\thepage}{S\arabic{page}}
\renewcommand{\thesection}{S\arabic{section}}
\renewcommand{\thetable}{S\arabic{table}}
\makeatletter

\renewcommand{\thesection}{\arabic{section}}
\renewcommand{\thesubsection}{\thesection.\arabic{subsection}}
\renewcommand{\thesubsubsection}{\thesubsection.\arabic{subsubsection}}

\vspace{0.3cm}
The Supplemental Material contains:
\begin{enumerate}
    \item Details on the analytical computations for annealed averages on the full unitary group \begin{itemize}
    \item Schur-Weyl duality and integration formulas for Haar unitaries
    \item Summary of the observables of interest and their replica functional formulation
    \end{itemize}
    \item Additional numerical results
    \begin{itemize}
        \item Self-averaging in encoding-decoding circuits
        \item Participation entropy for coherent errors and localization transitions
        \item Properties of the system in the code-space
        \item Dependence of participation and entanglement entropies on the R\'enyi index $q$
        \item Various choices of the initial logical state
        \item Numerical test for the code rate $r=1/4$
    \end{itemize}
    \item Robustness of the Main Text findings to inhomogeneities and additional sources of errors \begin{itemize}
        \item Analytical results for site-dependent disordered local errors
        \item Robustness of the transition for noisy decoders.
    \end{itemize}
\end{enumerate}
The Mathematica notebook for the symbolic manipulations, and the closed expression for the observables of interest are publicly available at~\cite{turkeshi_2023_10302870}.

\section{Details on the analytical computations for annealed averages on the full unitary group}
\subsection{Schur-Weyl duality and integration formulas for Haar unitaries}
This section reviews the properties of the averaged $k$-tensor product of Haar unitaries acting on $N$ qubits. (General references on the topic are~\cite{Roberts_2017,roth2018recoveringquantumgates,Gross2021}).  In the following, we denote the dimension of the Hilbert space of a single qubit by $d$, which for the systems of interest is given by $d=2$.
The core object is the averaged channel 
\begin{equation}
   \Phi^{(k)}_{\mathrm{Haar}}(O)= \mathbb{E}_{U\in \mathcal{U}(d^N)}\left[(U^\dagger)^{\otimes k} O U^{\otimes k} \right]\equiv \int_{\mathrm{Haar}} d\mu(U) (U^\dagger)^{\otimes k} O U^{\otimes k}\label{eq:swd}
\end{equation}
appearing multiple times in the Main Text. Here $\mathcal{U}(d^N)$ is the full unitary group on the Hilbert space $\mathbb{C}^{d N}$ with the Haar measure $\mu(U)$. 
The critical result simplifying the evaluation of~\eqref{eq:swd} is the Schur-Weyl duality. In a nutshell, this theorem states that any operator acting on $(\mathbb{C}^{d N})^{\otimes k}$ that commutes with $U^{\otimes k}$ for each unitary $U\in \mathcal{U}(d^N)$ is a linear combination of permutation operators. The set of all such operators is called the commutant of $\mathcal{U}(d^N)$.
We note that $\Phi^{(k)}_{\mathrm{Haar}}(O)$ is an element of the commutant. This follows from the group invariance of $\mu(U)$
\begin{equation}
    (V^\dagger)^{\otimes k} \Phi^{(k)}_{\mathrm{Haar}}(O) V^{\otimes k} = \int_{\mathrm{Haar}} d\mu(U) (V^\dagger)^{\otimes k}  (U^\dagger)^{\otimes k} O  U^{\otimes k}  V^{\otimes k} = \int_{\mathrm{Haar}} d\mu(\tilde{U})  (\tilde{U}^\dagger)^{\otimes k} O  {\tilde{U}}^{\otimes k} = \Phi^{(k)}_{\mathrm{Haar}}(O).
\end{equation}
Hence, we can write
\begin{equation}
    \Phi^{(k)}_{\mathrm{Haar}}(O) = \sum_{\pi\in \mathcal{S}_k} b_\pi(O) T_\pi,\label{eq:haar1}
\end{equation}
where $\mathcal{S}_k$ is the permutation group over $k$ elements and $T_\pi$ is its representation acting on $(\mathbb{C}^{d N})^{\otimes k}$. We recall that this is a tensor representation $T_\pi = \otimes_{i=1}^N t^{(i)}_\pi$, where $t^{(i)}_\pi$ is the permutation operation acting on the $i$-th site~\cite{Gross2021}. 
Playing with the above formula and with the Haar invariance, one can show that the coefficients $b_\pi(O)$ are given by
\begin{equation}
   b_\pi = \sum_{\sigma \in S_k} W_{\pi,\sigma} \mathrm{tr}\left( O T_\sigma\right),\label{eq:haar2}
\end{equation}
where $W_{\pi,\sigma}$ are the Weingarten symbols given by the inverse of the matrix $Q_{\sigma,\tau}\equiv \mathrm{tr}(T_\sigma T_\pi)=d^{N \#\mathrm{cycles}(\sigma\tau)}$. 
Denoting the projectors over the irreducible representations of $\mathcal{S}_k$ by $\Pi_\lambda$, their dimensions by $d_\lambda$ and their character functions by $\chi_\lambda$, the Weingarten symbols can be written as a sum over the integer partitions of $k$
\begin{equation}
    W_{\pi,\sigma} = \sum_{\lambda \vdash k} \frac{d_\mathcal{B}^2}{(k!)^2} \frac{\chi_\lambda(\pi\sigma)}{\mathrm{tr}(\Pi_\lambda)}.\label{eq:wein}
\end{equation}
Combining Eq.~\eqref{eq:haar1}, Eq.~\eqref{eq:haar2}, and Eq.~\eqref{eq:wein} we can find the averaged channel $\Phi^{(k)}_{\mathrm{Haar}}(O)$ for the operator $O$ of interest. 

For our operator of interest, $A_{U,\mathcal{E}}^{(2q)}$, we have to compute $b_\pi(\mathcal{E})$. The coherent channels is fixed by $\mathcal{E}(\rho)=K_{\vec{0}} \rho K_{\vec{0}}^\dagger$, with $K_{\vec{0}}=\prod_{j=1}^N \exp[-i \alpha \sigma^z_i/2]$ (see Main Text). 
Then, computing $b_\pi(\mathcal{E})$, hence $\mathcal{A}_\mathcal{E}^{(2q)}$, amounts to evaluate
\begin{equation}
    b_\pi(\mathcal{E}_\mathrm{coherent}) =   \sum_{\tau,\pi}W_{\tau,\pi}  \mathrm{tr}\left(t_\pi (\exp[-i \alpha \sigma^z_i]\otimes \exp[+i \alpha \sigma^z_i])^{\otimes q}\right)^N,
\end{equation}
which simplifies to a counting problem, using $\exp(-i \alpha \sigma^z/2)\exp(+i \alpha \sigma^z/2)=\openone_2$ and $\mathrm{tr}(\exp(-i\alpha s \sigma^z))= 2 \cos^{s}(\alpha/2)$.
For the incoherent case $\mathcal{E}=\mathcal{E}_1\circ\dots \circ \mathcal{E}_N$, with $\mathcal{E}_i(\rho) = \sum_{\mu_i=0}^3 K_{\mu_i,i}\rho K_{\mu_i,i}^\dagger $ (see Main Text) we introduce the projector $Q_{2}=\sum_{P=\openone_2,\sigma^x,\sigma^y,\sigma^z} P^{\otimes 2}/4$. Recalling that $T_\pi = \otimes_{i=1}^N t^{(i)}_\pi$, where $t^{(i)}_\pi$ is the permutation operation acting on the $i$-th site~\cite{Gross2021}, we have after some algebra
\begin{equation}
    b_\pi(\mathcal{E}_\mathrm{incoherent}) =   \sum_{\tau,\pi}W_{\tau,\pi}  \mathrm{tr}\left(t_\pi \left[(1-\lambda)\openone_4+ \lambda Q_{2} \right]^{\otimes q}\right)^N,
\end{equation}
where in the last step we dropped the dependence on $i$ due to each contribution being equivalent. Using that $P^2 = \openone{}$  and $\mathrm{tr}(P)=\delta_{P,\openone{}_2}$, again this computation reduces to a combinatorial problem. Equipped with the above expressions for $b_\pi(\mathcal{E})$, we can perform exactly the various computations, as detailed below.

\subsection{Replica functionals for the observables of interest}
Here, we report the formulae for the replica functionals discussed in the Main Text and specify their replica boundary conditions. We shall use basic properties of tensor products, such as $\langle \phi_1 |A| \phi_2\rangle \langle \psi_1|B|\psi_2\rangle = \langle \phi_1, \psi_1|A\otimes B|\phi_2,\psi_2\rangle$, and $AB\otimes CD = (A\otimes C)(B\otimes D)$. 
We recall from the Main Text that the error channels of interests are given by $\mathcal{E}(\rho)=\sum_{\vec{\mu}} K_{\vec{\mu}} \rho K_{\vec{\mu}}^\dagger$, with $\vec{\mu} = \vec{0}$ and $K_{\vec{0}}=\prod_{j=1}^N \exp[-i \alpha \sigma^z_i/2]$ for the coherent case, while 
$\vec{\mu}$ ranges over $4^N$ values for the depolarizing case. 
Beyond the observables of interest in the Main Text, we will also consider the participation entropy for the logical and code-space decoded states, namely~\cite{luitz2014shannonrenyientropy,sierant2022universalbehaviorbeyond}
\begin{align}
    S^\mathrm{part,X}_{q} &=  \frac{1}{1-q}\log_2\left(\sum_{m=0}^{2^{k}-1} \langle m_X | \rho_X |m_X\rangle^{q}\right), \\
    S^\mathrm{part}_{q} &=  \frac{1}{1-q}\log_2\left(\sum_{m=0}^{2^{N}-1} \langle m | \rho |m\rangle^{q}\right), 
\end{align}
for which we derive closed expressions for $q=2$ and $q=3$, and the leading order for $q\ge 4$. 
To keep the presentation concise, we will present the step-by-step derivation only for the fidelity between the initial (pure) and final (in general mixed) logical state. The other derivations proceed in a similar fashion, and we simply report the final results, stressing their replica boundary operators.
The fidelity is given by $F=\langle \psi_X|\rho_X|\psi_X\rangle$, cf.~Eq.~(3) in the Main Text.
Then, the numerator contribution is
\begin{equation}
\begin{split}
   \mathrm{numerator}(\Phi_F^{(2)})\equiv\sum_{{\vec{\mu}}} &\langle (\psi_X,0_{\bar{X}}) | U^\dagger K_{\vec{\mu}}^\dagger U |(\psi_X,0_{\bar{X}})\rangle\langle (\psi_X,0_{\bar{X}})|U^\dagger  K_{\vec{\mu}} U |(\psi_X,0_{\bar{X}})\rangle= \\&= \langle (\psi_X,0_{\bar{X}})|^{\otimes 2} A^{(2)}_{U,\mathcal{E}} |(\psi_X,0_{\bar{X}})\rangle^{\otimes 2}=\mathrm{tr}(\mathcal{B}_\mathrm{num}^F(2) A^{(2)}_{U,\mathcal{E}}),
\end{split}
\end{equation}
recovering $\mathcal{B}^F_\mathrm{num}$ in the Main Text and $A_{U,\mathcal{E}}^{(2)}$ defined in Eq. (2) of the Main Text, with $q=1$. 
Similarly, for the denominator 
\begin{equation}
\begin{split}
   \mathrm{denominator}(\Phi_F^{(2)})\equiv & \sum_{\vec{\mu}} \langle (\psi_X,0_{\bar{X}}) | U^\dagger K_{\vec{\mu}}^\dagger U \openone_X\otimes|0_{\bar{X}})\rangle\langle 0_{\bar{X}}|U^\dagger K_{\vec{\mu}} U |(\psi_X,0_{\bar{X}})\rangle =\\
    &=\sum_{\vec{\mu}} \sum_{x=0}^{2^k-1}\langle (\psi_X,0_{\bar{X}}) | U^\dagger K_{\vec{\mu}}^\dagger U |(x,0_{\bar{X}}))\rangle\langle (x,0_{\bar{X}})|U^\dagger K_{\vec{\mu}} U |(\psi_X,0_{\bar{X}})\rangle = \mathrm{tr}(\mathcal{B}_\mathrm{den}^F(2) A^{(2)}_{U,\mathcal{E}}),
\end{split}
\end{equation}
where we have used the resolution of the identity in the first step and read out $\mathcal{B}_\mathrm{den}^F$ described in the Main Text. We remark that the order of the states in the tensor products is crucial for the correct computation (e.g., note the inverted order between $(\psi_X,0_{\bar{X}})$ and $(x,0_{\bar{X}})$, resulting in a swap permutation $T_{(12)}$ as illustrated in the Main Text). Finally, we note that the calculation of the annealed averages for $F$ is relatively easy, as it requires the Weingarten calculus over the symmetry group $\mathcal{S}_2$.

For the \emph{participation entropy} in the logical space $\rho_X$, we have 
\begin{equation}
\begin{split}
    S_q^\mathrm{part} &= \frac{1}{1-q}\log_2\left[\frac{\mathrm{tr}( A^{(2q)}_{U,\mathcal{E}}\mathcal{B}_{\mathrm{num}}^\mathrm{part,X}(2q)) }{ \mathrm{tr}(A^{(2q)}_{U,\mathcal{E}}\mathcal{B}_{\mathrm{den}}^\mathrm{part,X}(2q) )}\right], \\
    \mathcal{B}_\mathrm{num}^\mathrm{part,X} &= \sum_{m_X=0}^{2^k-1} \rho_0^{\otimes q}\otimes |m_X, 0_{\bar{X}}\rangle\langle m_X, 0_{\bar{X}}|^{\otimes q} T_{(1,2,\dots,2q-1,2q)}\\
    \mathcal{B}_\mathrm{den}^\mathrm{part,X} &= \rho_0^{\otimes q}\otimes (\openone_X\otimes | 0_{\bar{X}}\rangle\langle0_{\bar{X}}|)^{\otimes q} T_{(1,2,\dots,2q-1,2q)}
\end{split}
\end{equation}
where the cycle permutation $(1,2,\dots,2q-1,2q)$ rotate all the replicas. Thus, the annealed averages $\tilde{S}_q^\mathrm{part}$ require Haar integrals and Weingarten calculus of order $2q$, amenable for $q=2$ and $q=3$ in a closed expression, and for $q=4$ at leading order in Hilbert space dimension.
As we present in the Sec. "Additional numerical results"  of this Supplemental Material, our analytic results quantitatively match the numerical quench averages. In particular, we find that a localization transition occurs, with $D_q=0$ below the critical error strength, and $D_q>0$ otherwise. Similarly to the entropy in the Main Text, $D_q$  saturates to $1$ at a value of error strength that depends non-trivially on $q$, highlighting again the multifractal features of the system~\cite{luitz2014universalbehaviorbeyond,sierant2022universalbehaviorbeyond}.

Similarly, for the entropy (entanglement or thermodynamic depending on the 
choice of the subsystems $X_1$ and $X_2$) 
of $\rho_X$ as discussed in the main Main Text, we find
\begin{equation}
\begin{split}
    S_q &= \frac{1}{1-q}\log_2\left[\frac{\mathrm{tr}( A^{(2q)}_{U,\mathcal{E}}\mathcal{B}_{\mathrm{num}}^\mathrm{ent,X}(2q)) }{ \mathrm{tr}(A^{(2q)}_{U,\mathcal{E}}\mathcal{B}_{\mathrm{den}}^\mathrm{ent,X}(2q) )}\right], \\
    \mathcal{B}_\mathrm{num}^\mathrm{ent,X} &=  \rho_0^{\otimes q}\otimes (\openone_X\otimes | 0_{\bar{X}}\rangle\langle0_{\bar{X}}|)^{\otimes q} \left(\bigotimes_{i\in X_1} t^{(i)}_{(1,3,\dots,2q-1)}\right)\otimes \left(\bigotimes_{j\in X_2} t^{(j)}_{()}\right) T_{(1,2,\dots,2q-1,2q)}\\
    \mathcal{B}_\mathrm{den}^\mathrm{ent,X} &= \mathcal{B}_\mathrm{den}^\mathrm{part,X}
\end{split}
\end{equation}
where $t^{(i)}_\pi$ are the permutation representation acting on the replicated sites, $(1,3,\dots,2q-1)$ is a cyclic permutation ensuring the computation of $\rho_{X_1}^q$, and $()$ is the trivial permutation. Notice the fact that the denominators coincide with that of the participation entropy. 
To analyze the properties of the code-space state, we consider the state ${\rho = U^\dagger \mathcal{E}(U \rho_0 U^\dagger )U}$ before the projection of the ancilla qubits. This results in 
\begin{equation}
\begin{split}
        S_q^\mathrm{part} &=  \frac{1}{1-q}\log_2\left[\sum_{m=0}^{2^N-1}\left(\langle m|\rho| m\rangle\right)^q\right] =\frac{1}{1-q}\log_2\left[\mathrm{tr}(A^{(2q)}_{U,\mathcal{E}}\mathcal{B}^\mathrm{part}(2q) )\right]\\
        \mathcal{B}^\mathrm{part}(2q) &= \sum_{m=0}^{2^N-1} \rho_0^{\otimes q}\otimes |m\rangle\langle m|^{\otimes q} T_{(1,2,\dots,2q-1,2q)}
\end{split}
\end{equation}
while for the entropy (see Main Text) we have 
\begin{equation}
\begin{split}
        S_q &=  \frac{1}{1-q}\log_2\left[\mathrm{tr}(A^{(2q)}_{U,\mathcal{E}}\mathcal{B}^\mathrm{ent}(2q) )\right]\\ 
        \mathcal{B}^\mathrm{ent}(2q) & = \rho_0^{\otimes q}\otimes (\openone_{X\cup\bar{X}})^{\otimes q} \left(\bigotimes_{i\in X} t^{(i)}_{(1,3,\dots,2q-1)}\right)\otimes \left(\bigotimes_{j\in \bar{X}} t^{(j)}_{()}\right)T_{(1,2,\dots,2q-1,2q)}.
\end{split}
\end{equation}
As for the logical state entanglement entropy, calculating the exact annealed average $\tilde{S}^\mathrm{ent}_q$ requires  Weingarten integrals of order $2q$. 
The annealed averaged values are however involved and lengthy. We present them in an operationally accessible fashion in the open-access folder~\cite{turkeshi_2023_10302870}.

\section{Additional numerical details}

\begin{figure}[ht]
    \centering
    \includegraphics[width=\columnwidth]{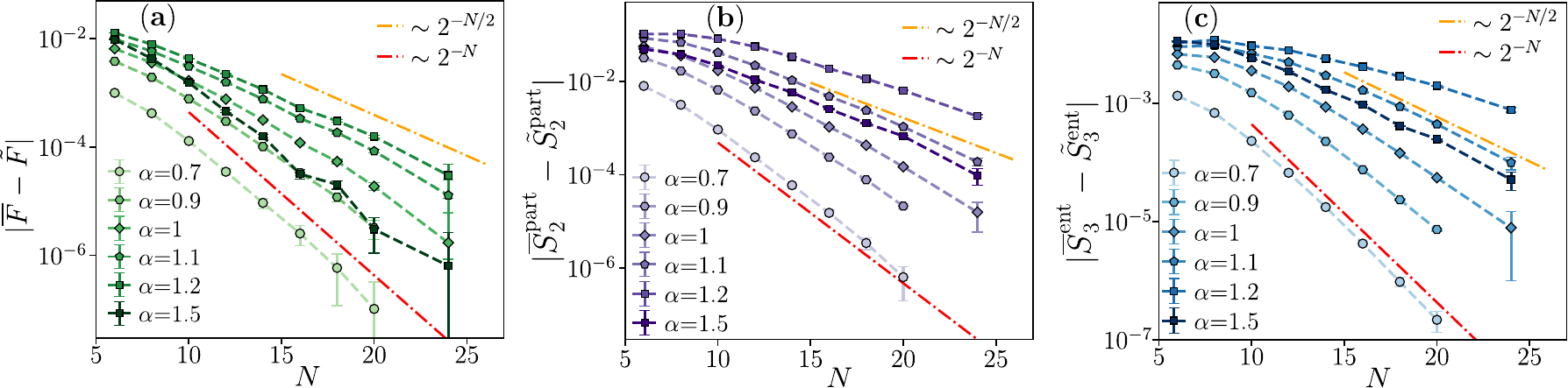}
    \caption{Difference between the quenched and annealed averages of the fidelity $F$, participation entropy $S^{\mathrm{part}}_q$ and entanglement entropy $S^{\mathrm{ent}}_q$ of the logical decoded state $\rho_X$ for various error strengths $\alpha$ as function of the system size $N$.  The code rate is fixed as $r=1/2$ and subsystem used in evaluation of the entanglement entropy is composed of $|X_1|=N/4$ qubits. The dash-dotted lines illustrate the observed exponential decays $e^{-\kappa N}$. }
    \label{fig:self}
\end{figure}

\subsection{Self-averaging in the encoding-decoding circuits}
The self-averaging of the fidelity and entropies in the encoding-decoding circuits provides insights into the properties of the error-resilience phase transitions. Here, we recall the notions of the quenched and annealed averages introduced in the Main Text. We argue that the former averages follow the definitions in a more "natural" fashion, while the latter are more straightforward from the perspective of the exact analytical calculations. We demonstrate the self-averaging properties that allow us to interchangeably use the quenched and the annealed averages in the limit of large number qubits, $N \gg 1$.

A single realization of the encoding-decoding circuit is specified by the encoding unitary $U$ selected with Haar measure from the unitary group $\mathcal U(2^N)$ and strength $\alpha$ (or $\lambda$) for coherent (incoherent) errors. To obtain the final results for a quantity of interest $\Lambda$ at a given $\alpha$ (or $\lambda$), we consider:
\begin{itemize}
    \item quenched averages, $\overline \Lambda
\equiv \mathbb{E}_{U}\left[\Phi^{(2q)}_\Lambda({A}_{U,\mathcal{E}}^{(2 q)})\right]$, where $\Phi^{(2q)}_\Lambda({A}_{U,\mathcal{E}}^{(2 q)})$ is the value of $\Lambda$ for a given realization of the circuit; once $\Lambda$ is evaluated, the result is averaged over the realizations of the circuit, i.e., the average over the unitary group, $\mathbb{E}_{U}(.)$, is taken;
    \item annealed averages, $\tilde \Lambda
\equiv \Phi^{(2q)}_\Lambda\left[\mathbb{E}_{U}\left({A}_{U,\mathcal{E}}^{(2 q)}\right)\right]$, for which the first step is to average the operator ${A}_{U,\mathcal{E}}^{(2 q)}$ over the unitary group, and the second step is to evaluate the functional $\Phi^{(2q)}_\Lambda$ of the averaged operator   $\mathcal{A}_\mathcal{E}^{(2q)} =\mathbb{E}_{U}\left({A}_{U,\mathcal{E}}^{(2 q)} \right)$. 
\end{itemize}
Let us consider the fidelity, $F$, as an example to illustrate the differences between the two types of averages (the reasoning for the participation, entanglement and thermodynamic entropies is analogous). The fidelity can be rewritten as the following functional over the replica space:
\begin{equation}
 F= \Phi_{{F}}^{(2)}({A}_{U,\mathcal{E}}^{(2)}) 
 =  \frac{\langle (\psi_X,0_{\bar{X}}),(\psi_X,0_{\bar{X}})|{A}_{U,\mathcal{E}}^{(2)}| (\psi_X,0_{\bar{X}}),(\psi_X,0_{\bar{X}})\rangle}{
 \sum_{x=0}^{2^k-1}
 \langle (\psi_X,0_{\bar{X}}),(x,0_{\bar{X}})|{A}_{U,\mathcal{E}}^{(2)}| (x,0_{\bar{X}}),(\psi_X,0_{\bar{X}})\rangle},
\end{equation}
which is identical to Eq.~(3) of the Main Text. A "natural" procedure is to calculate $F$ for each realization of the encoding-decoding circuit and to average the results, which yields the quenched average $\overline F$. We employ this procedure in our numerical computations. Alternatively, we may follow the annealed average prescription and evaluate
\begin{equation}
\tilde{F} \equiv \Phi_{{F}}^{(2)}(  \mathcal{A}_{\mathcal{E}}^{(2)}  ) 
 =  \frac{\langle (\psi_X,0_{\bar{X}}),(\psi_X,0_{\bar{X}})|\mathcal{A}_{\mathcal{E}}^{(2)}  | (\psi_X,0_{\bar{X}}),(\psi_X,0_{\bar{X}})\rangle}{
 \sum_{x=0}^{2^k-1}
 \langle (\psi_X,0_{\bar{X}}),(x,0_{\bar{X}})| \mathcal{A}_{\mathcal{E}}^{(2)} | (x,0_{\bar{X}}),(\psi_X,0_{\bar{X}})\rangle}.\label{eq:tildeDEL}
\end{equation}
To that end, we start with the calculation of the operator $\mathcal{A}_{\mathcal{E}}^{(2)}$ averaged over the unitary group $\mathcal U(2^N)$ and then compute the ratio of the matrix elements. Since both the numerator and denominator are linear in ${A}_U^{(2)}$, we can independently calculate the numerator 
\begin{equation}
    m_2 = \langle (\psi_X,0_{\bar{X}}),(\psi_X,0_{\bar{X}})|{A}_{U,\mathcal{E}}^{(2)}  | (\psi_X,0_{\bar{X}}),(\psi_X,0_{\bar{X}})\rangle
    \label{eq:num}
\end{equation} 
and denominator
\begin{equation}
    p_2 = \sum_{x=0}^{2^k-1}
 \langle (\psi_X,0_{\bar{X}}),(x,0_{\bar{X}})| {A}_{U,\mathcal{E}}^{(2)} | (x,0_{\bar{X}}),(\psi_X,0_{\bar{X}})\rangle,\label{eq:denum}
\end{equation} 
for each circuit realization, and then calculate the averages  $\mathbb{E}_{U}(m_2)$, $\mathbb{E}_{U}(p_2)$ to arrive at 
\begin{equation}
\tilde{F} =  \frac{\mathbb{E}_{U}(m_2)}{\mathbb{E}_{U}(p_2)}.\label{eq:tildeDEL2}
\end{equation}
Hence, the calculation of the annealed averages amounts to taking independent averages of the numerators and denominators in the involved functionals. This indicates a straightforward path to calculate the annealed averages in numerical computations and explains why the annealed averages are simpler to tackle with the exact analytic approach.

To illustrate these arguments, we focus on the coherent errors. Similar results holds also for the incoherent case, and are not reported here. 
In Fig.~\ref{fig:self}, we compare annealed and quenched averages of the fidelity and of the participation/entanglement entropies. All of the presented quantities were calculated numerically (we have checked that the exact analytical expressions for the annealed averages agree with numerical results within the estimated error bars). The panels (a),(b) and (c) of Fig.~\ref{fig:self} show that the difference between the annealed and the quenched averages of $F$, $S^{\mathrm{part}}_q$ and $S^{\mathrm{ent}}_q$ decreases exponentially $\sim e^{-\kappa N}$ with system size $N$, where $\kappa$ is a rate that depends on the considered quantity and the error strength $\alpha$. This indicates that the annealed and quenched averages behave in the same way in the $N\gg1$ limit, and both of them can be used in investigations of the EPP to EVP transition in our setup.

\begin{figure}
    \centering
    \includegraphics[width=0.66\columnwidth]{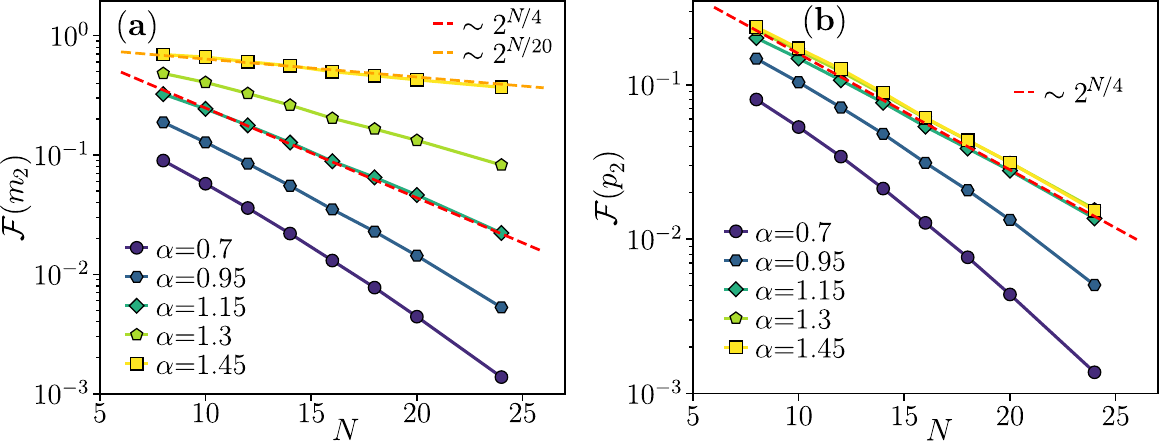}
    \caption{Mechanism of the self averaging for the fidelity $F$. Numerically calculated realization-to-realization fluctuations \eqref{eq:fluct1} and \eqref{eq:fluct2} of the numerator and denominator involved in the calculation of the fidelity (cf.~\eqref{eq:tildeDEL2}) are shown by the lines with markers for various error strengths $\alpha$. The dashed lines illustrate exponential decay $e^{-\gamma N}$ with the number of qubits $N$. The code rate is fixed as $r=1/2$.}
    \label{fig:self2}
\end{figure}
\begin{figure}[ht]
    \centering
    \includegraphics[width=\columnwidth]{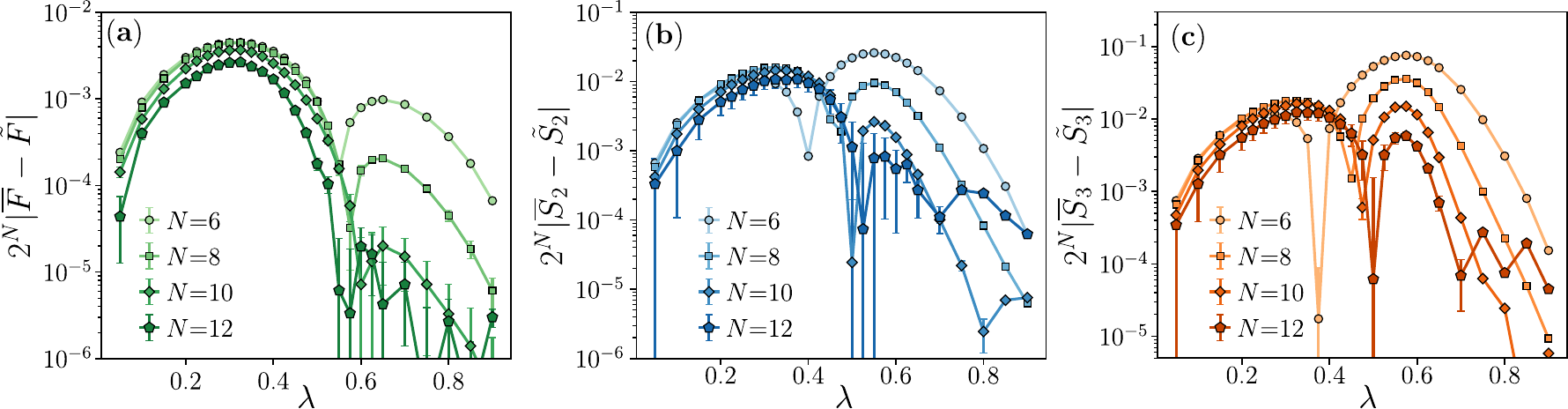}
    \caption{Self averaging for the depolarizing noise. Rescaled difference between the quenched and annealed averages of the fidelity $F$, and thermodynamic entropies $S_q$ of the logical state $\rho_X$ as function of the error strength $\lambda$. The rescaling factor is $2^N$ indicating (super)exponential with system size $N$ decay  of the difference between quenched and annealed averages. The code rate is fixed as $r=1/2$. }
    \label{fig:selfdepol}
\end{figure}

To understand the reason underlying the self-averaging, we consider again the fidelity $F$. The numerator $m_2$~\eqref{eq:num} and the denumerator $p_2$~\eqref{eq:denum} become random variables dependent on the choice of the unitary $U$ that specifies the circuit. We quantify their relative realization-to-realization fluctuations through the ratios of their standard deviations to their average values
\begin{equation}
\mathcal{F}(m_2)= \frac{\sqrt{\mathbb{E}_{U}\left[m_2^2 - (\mathbb{E}_{U}(m_2))^2\right]}}{\mathbb{E}_{U}(m_2)}
\label{eq:fluct1}
\end{equation}
and
\begin{equation}
\mathcal{F}(p_2)= \frac{\sqrt{\mathbb{E}_{U}\left[p_2^2 - (\mathbb{E}_{U}(p_2))^2\right]}}{\mathbb{E}_{U}(p_2)}.
\label{eq:fluct2}
\end{equation}
Calculating numerically the fluctuations $\mathcal{F}(p_2)$ and $\mathcal{F}(m_2)$, we obtain results shown in Fig.~\ref{fig:self2}. The results indicate that the relative realization-to-realization fluctuations decay exponentially with the number of qubits: $\mathcal{F}(p_2) \sim e^{-\gamma N}$, where $\gamma$ is a rate that depends on the value of $\alpha$. The same behavior is observed for $\mathcal{F}(m_2)$. This exponential damping of the realization-to-realization fluctuations shows why the calculation of the quenched average (the whole quantity is averaged) and the annealed average (the numerator and denominator are averaged independently) yield the same results. Indeed, for a fixed circuit realization, we have $m_2 = (1 + r_1 \mathcal F(m_2) ) \mathbb{E}_{U}(m_2)$ and $p_2 = (1 + r_2 \mathcal F(p_2) ) \mathbb{E}_{U}(p_2)$, where $r_1$ and $r_2$ are random numbers of the order of unity, which yields
\begin{equation}
\frac{m_2}{p_2} = \frac{\mathbb{E}_{U} (m_2) }{\mathbb{E}_{U} (p_2) } \left(1 - r_2\mathcal{F}(p_2)+r_1\mathcal{F}(m_2) \right) + O(r_2^2) = \tilde F + O( e^{-\gamma N}).\label{eq:selfav}
\end{equation}
Averaging this equation over the unitary group $\mathcal U(2^N)$, we obtain the quenched average $\overline F$ on the left-hand side. Hence, the difference between the quenched average, $\overline F$, and the annealed average, $\tilde F$, is exponentially small in the system size $N$. Moreover, Eq.~\ref{eq:selfav} demonstrates that $F$ evaluated for a single circuit realization approximates the quenched and annealed averages over the unitary group with an error that decreases exponentially with $N$, demonstrating the self-averaging of the fidelity. Analogous reasoning applies to other considered quantities, and also to the case of the depolarizing noise as shown in Fig.~\ref{fig:selfdepol}.

\begin{figure*}
    \centering
    \includegraphics[width=.37\textwidth]{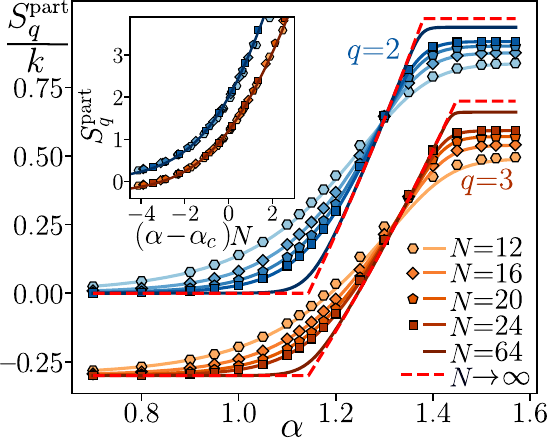}
    \caption{Error-resilience phase transition at $r=k/N=1/2$. 
     Participation entropy $S^{\mathrm{part}}_q$  in the logical space for $q=2$ and $q=3$ (shifted by a constant).
    Markers and solid lines correspond to the numerical data for the quenched averages and analytic expressions for the annealed averages.
    The error bars are smaller than the marker size.
}
    \label{fig:partent}
\end{figure*}

\subsection{Participation entropy}
Throughout this section, we focus on the coherent errors. 
Similar results, albeit not reported here, hold also for the incoherent case (see~\cite{turkeshi_2023_10302870}).
The methods highlighted above allow us to compute in a closed form $\tilde{S}^\mathrm{part}_q$ for $q=2$ and $q=3$. The exact expressions are lengthy and given for arbitrary $N$, $k$, and $\alpha$ in~\cite{turkeshi_2023_10302870}. Here, we report the leading scaling $\tilde{S}^\mathrm{part}_q=D_q k$ at $N\to \infty$, with the multifractal dimension given by 
\begin{equation}
    D_q = \begin{cases} 0,& \alpha<\alpha_c(r)\\
    \frac{q \left((r-1) -\log_2 \left(\cos^2 \left(\alpha/2\right)\right)\right)}{(q-1) r }, & \alpha_c(r)<\alpha<\alpha^\mathrm{part}_X(r),\\
    1,&\text{otherwise}.
    \end{cases} \label{eq:partent}
\end{equation}
This result reveals that the error-resilience phase transition at $\alpha $$=$$\alpha_c$ manifests itself as a transition between localization ($D_q=0$) and delocalization ($D_q>0$) in the computational basis of the logical space. Interestingly, the error strength $\alpha^\mathrm{part}_X(r) $ at which the state becomes fully extended, i.e. when $D_q=1$, depends non-trivially on the R\'enyi index, $\alpha^\mathrm{part}_X(r) = 2 \arccos \left(2^{-{(q-r)}/{(2 q)}}\right)$. Hence, our results demonstrate the existence of an extended interval of error strength within the EVP, $\alpha_c(r)<\alpha<\alpha^\mathrm{part}_X(r)$, in which the logical state is multifractal.
While the presence of multifractal and delocalized states is analogous to earlier observations for quantum many-body systems~\cite{stephan2009,luitz2014universalbehaviorbeyond,lindinger2019manybodymultifractality,mace2019multifractal,Pietracaprina2021, roy2022hilbertspace,
sierant2022universalbehaviorbeyond}, the Hilbert space localization appears to be a unique feature of EPP. 

In Fig.~\ref{fig:partent}, we show that our exact analytic expressions for the annealed average of the participation entropy $\tilde{S}^\mathrm{part}_q$ are in excellent agreement with the numerically computed quenched averages $\overline{S}^\mathrm{part}_q$ already for $N=12$ for $q=2$ and $N=16$ for $q=3$.
We also observe that $\tilde{S}^\mathrm{part}_q$ approaches the thermodynamic limit result~\eqref{eq:partent} with increasing $N$ and that the participation entropy follows a critical behavior with exponent $\nu=1$ at $\alpha=\alpha_c$.

\begin{figure}
    \centering
    \includegraphics[width=0.5\columnwidth]{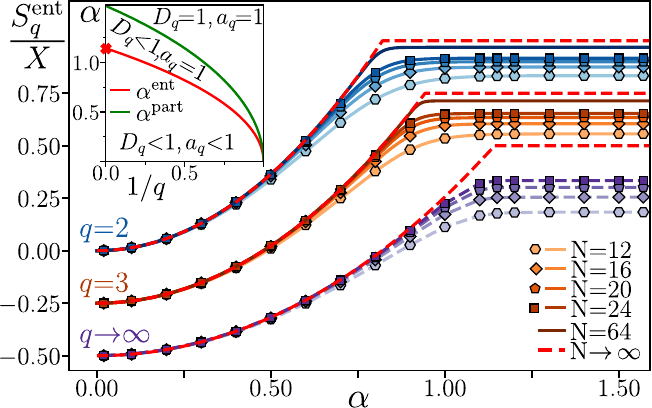}
    \caption{Entanglement entropy $S^{\mathrm{ent}}_q$ of the code-space state $\rho$ for the  bipartition $X \cup \overline X$, and $q=2,3,\infty$ (the $q=2,\infty$ results are shifted down respectively by $0.25$ and $0.5$). Markers and solid lines correspond, respectively, to the numerical data for the quenched averages $\overline S^{\mathrm{ent}}_q$ and exact analytic expressions for the annealed average $\tilde S^{\mathrm{ent}}_q$  for $q=2,3$, while the purple dashed lines for $q=\infty$ are to guide the eye. The red dashed lines correspond to the $N\to\infty$ behavior~\eqref{eq:code2}. The inset shows $\alpha^\mathrm{part}(q)$ and $\alpha^\mathrm{ent}(1/2,q)$ as a function of $1/q$. }
    \label{fig:codespace}
\end{figure}

\subsection{Properties of the system in code-space.}
This section focuses on the coherent error model. Similar results holds also for the incoherent case, see analytic formulae in~\cite{turkeshi_2023_10302870}.
We investigate the properties of the code-space state $\rho$ before the projection on the ancilla qubits, cf. Main Text. We calculate the R\'enyi participation and entanglement entropies \eqref{eq:defobs1} of $\rho$ (we recall that in the considered case $\rho$ is pure). For calculating $S^{\mathrm{ent}}_q$, we choose a bipartition of the full system in the logical, $X$, and ancilla, $|X|$, subsystems. The annealed averages $\tilde{S}^{\mathrm{ent}}_q$ are computed in closed form for $q=2, 3$, cf.~\cite{turkeshi_2023_10302870}. The leading behavior at large $N$ and at any $q>1$ is $\tilde{S}_q^\mathrm{ent} = a_q |X|$, with
\begin{equation}
    a_q = \begin{cases} 
    \frac{\log_2\left(\cos ^{(2 q)/r}\left(\alpha/2\right)\right)}{1-q},& \alpha<\alpha^\mathrm{ent}(r,q)\\
        1,&\text{otherwise}. 
    \end{cases}\label{eq:code2}
\end{equation}
This shows that the code-space state $\rho$ undergoes a volume-to-volume law entanglement transition at error strength $\alpha^\mathrm{ent}(r,q) = 2 \arccos \left(2^{\frac{r}{2 q}-\frac{r}{2}}\right)$.
We perform analogous calculations for the participation entropy of $|\Psi^U\rangle$, obtaining the exact results for the annealed averages for $q=2,3$ and finding that the leading behavior for $N\to\infty$ is $\tilde{S}_q^\mathrm{part}=D_q N$ with multifractal dimension given by
\begin{equation}
    D_q = \begin{cases} 
    \frac{\log_2\left(\cos ^{(2 q)}\left(\alpha/2\right)\right)}{1-q},& \alpha<\alpha^\mathrm{part}(q)\\
        1,&\text{otherwise}.
    \end{cases}\label{eq:code}
\end{equation}
Hence, the code-space state becomes fully delocalized for $\alpha >\alpha^\mathrm{part}(q)=2\arccos \left(2^{-(q-1) r/(2q)}\right)$. Both  $\alpha^\mathrm{ent}(1/2,q)$ as well as $\alpha^\mathrm{part}(q)$ depend non-trivially on $q$, as illustrated in the inset of Fig.~\ref{fig:codespace}, highlighting the multifractal behavior of the code-space state $\rho$. Moreover, $\lim_{q\to\infty} \alpha^\mathrm{ent}(r,q) = \alpha_c(r)$, signaling a link between the properties of the state in the code-space and the EPP to EVP transition in the logical space. Finally, we note that the comparison between quenched and annealed averages of entanglement entropy, shown in the onset of Fig.~\ref{fig:codespace}, demonstrates quantitative agreement between the two types of averages.

\subsection{Dependence of participation and entanglement entropies on the R\'enyi index $q$}

\begin{figure}[ht]
    \centering
    \includegraphics[width=0.7\columnwidth]{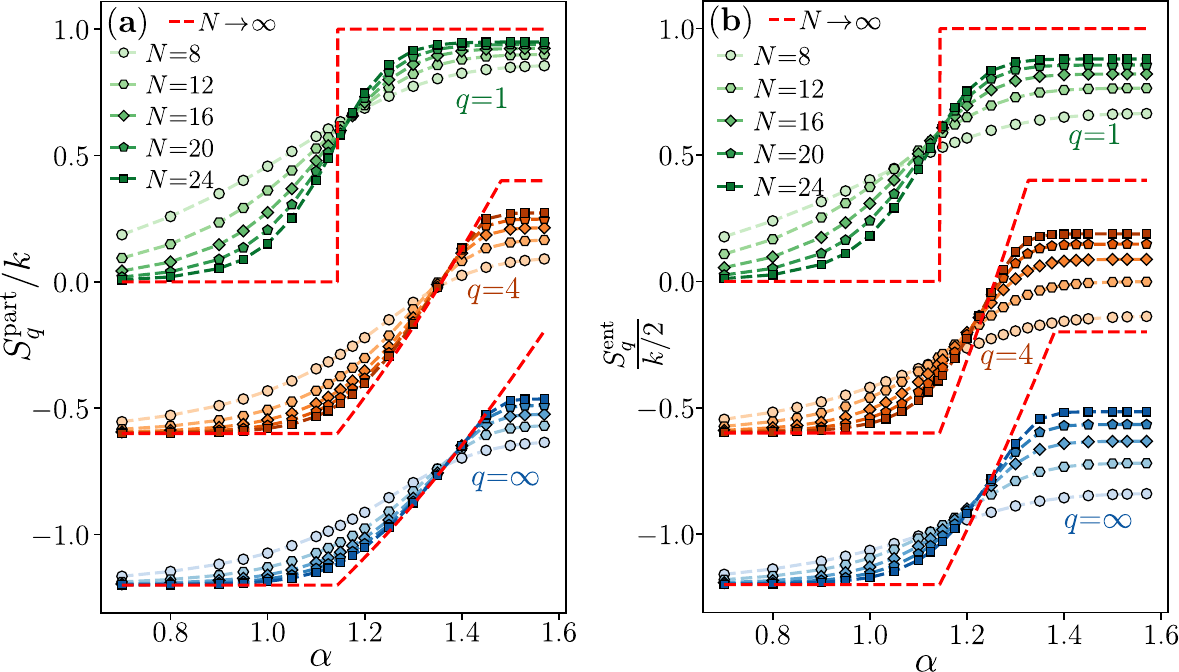}
    \caption{
    Error-resilience phase transition at $r=k/N=1/2$: R\'enyi index $q$ dependence of the participation entropy $S^{\mathrm{part}}_q$ and the entanglement entropy $S^{\mathrm{ent}}_q$ (the subsystem size is $|X_1|=k/2$) of the logical state $\ket{\psi^U_X}$ is shown respectively in panels (a) and (b). The markers show the numerically calculated annealed averages for $q=1,4,\infty$, while the red dashed lines show the leading behavior in the $N \to \infty$ limit, given by Eq.~(8) and Eq.~(9) of the Main Text. The data for $q=4$ and $q=\infty$ are shifted downwards by $0.6$ and $1.2$ for the clarity of presentation.
    The error bars are smaller than the marker size.}
    \label{fig:highQ}
\end{figure}

Our exact analytical calculations for the participation and entanglement entropies result in formulas valid for $q=2,3$ for arbitrary system size, code rate, subsystem size, and error strength $\alpha$. The observed behavior of the leading term in the expressions for $S^{\mathrm{part}}_q$ and $S^{\mathrm{ent}}_q$ of the logical state at $q=2,3$, and the leading order estimates for $q\ge 4$,  allowed us to conjecture that the leading of the Main Text are valid for any $q \geq 1$. A comparison, for coherent errors, of those expressions with the numerically calculated $S^{\mathrm{part}}_q$ and $S^{\mathrm{ent}}_q$ shown in Fig.~\ref{fig:highQ} indicates that those formulas are indeed valid for any $q \geq 1$. Similar results hold also for the incoherent error case.

\subsection{Various choices of the initial logical state}

\begin{figure}[ht]
    \centering
    \includegraphics[width=1\columnwidth]{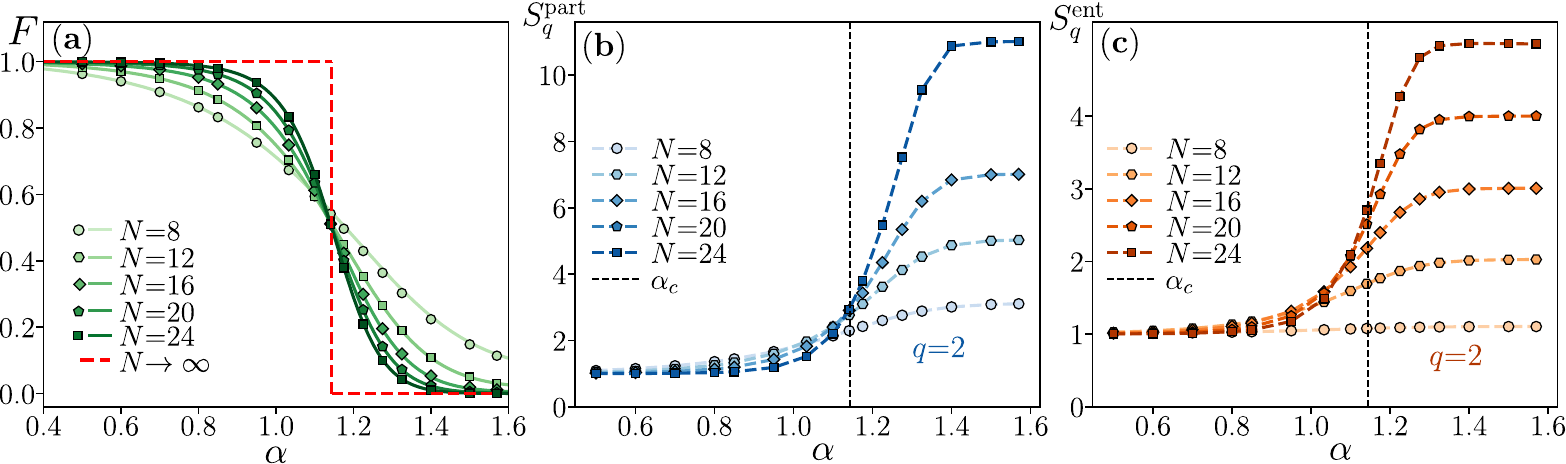}
    \caption{Error-resilience phase transition at a code rate $r=k/N=1/2$ with a GHZ state $\ket{\psi_{GHZ}}$ as the initial state; (a) the fidelity $F$, (b) the participation entropy $S^{\mathrm{part}}_q$, (c) the entanglement entropy $S^{\mathrm{ent}}_q$ (for subsystem size $|X_1|=N/4$) of the decoded logical state $\ket{\psi^U_X}$. The EPP to EVP transition is observed similarly as for the initial state $\ket{(0_X, 0_{\bar X} )}$, but $S^{\mathrm{part}}_q$ and  $S^{\mathrm{ent}}_q$ admit non-vanishing values in the EPP phase with $\ket{\psi_{GHZ}}$ as the initial state.
    The error bars are smaller than the marker size.}
    \label{fig:ghz}
\end{figure}

Throughout this section, we focus on coherent errors; analogous results, not reported, hold also for the depolarizing noise.
In the Main Text, we have fixed the initial state to be $\ket{\psi} = \ket{(0_X, 0_{\bar X})}$, where $ \ket{0_X} = \ket{0}^{\otimes k}$ is the state of the logical qubits and  $ \ket{0_{\bar X}} = \ket{0}^{\otimes(N-k)}$ is the state of the ancilla qubits. The EPP to EVP transition is independent of the choice of the initial logical state $\ket{\psi_X}$. However, the considered quantities may behave differently for various choices of $\ket{\psi_X}$.

The average fidelity, $F=\langle \psi_X |\rho_X|\psi_X\rangle$, is fully independent of the choice of the initial logical state $\ket{\psi_X}$. Indeed, by the translational invariance of the Haar measure, we can separate arbitrary fixed unitary operation from the encoding unitary: $U \to  U U_0$ and use $U_0$ to transform any initial state $\ket{\psi_X}$ to $\ket{0_X}$. 

The average entanglement entropy remains the same if $U_0$ comprises on-site unitaries: $U_0 = \prod_{i=1}^N U_i$. Hence, an initial state that is a product of spins pointing in random directions yields the same average $S^{\mathrm{ent}}_q$ as a function of $\alpha$ (as we checked numerically, data not shown). However, we note that the same is not valid for the participation entropy $S^{\mathrm{part}}_q$: if we started from a product state of spins polarized along the $X$ direction, we would observe $S^{\mathrm{part}}_q\propto N$ at any $\alpha$; in such a case it could be still possible to observe the EPP to EVP transition on the level of sub-leading terms, cf.~\cite{sierant2022universalbehaviorbeyond}, but this is beyond the present work. 

To test the above conclusions in a non-trivial manner, we initialize the logical qubits in a GHZ state:
\begin{equation}
\ket{\psi_{X,GHZ}} = \frac{1}{\sqrt{2}} ( \ket{0}^{\otimes k}+\ket{1}^{\otimes k}),
\end{equation}
which is an entangled state with an area-law entanglement entropy and a system size-independent participation entropy. Initializing the ancilla qubits in the state $\ket{0_{\bar X}}$, so that the full initial state is $\ket{\psi_{GHZ}}=\ket{(\psi_{X,GHZ},0_{\bar X})}$, we obtain results presented in Fig.~\ref{fig:ghz}. 
We observe the EPP to EVP transition. The fidelity $F$, consistently with our expectations, follows exactly the results of the Main Text for the initial state $\ket{(0_X,0_{\bar X})}$, and faithfully reproduces the exact analytical expressions for the quenched averages. Notably, the behavior of the participation entropy and the entanglement entropy in the EPP, for $\alpha < \alpha_c$, is different: both quantities approach, with increasing $N$, their GHZ state values, $S^{\mathrm{part}}_q=1$ and $S^{\mathrm{ent}}_q=1$. Hence, the error-resilience transition is accompanied by an area-to-volume law transition in the entanglement/participation entropy also with $\ket{\psi_{GHZ}}$ as the initial state. However,  for $\ket{\psi_{GHZ}}$, non-zero values of $S^{\mathrm{part}}_q$ and $S^{\mathrm{ent}}_q$ are found in the EPP, as ensured by the error-protecting properties of this phase.

\subsection{Numerical test for the code rate $r=1/4$.}
\begin{figure}
    \centering
    \includegraphics[width=1\columnwidth]{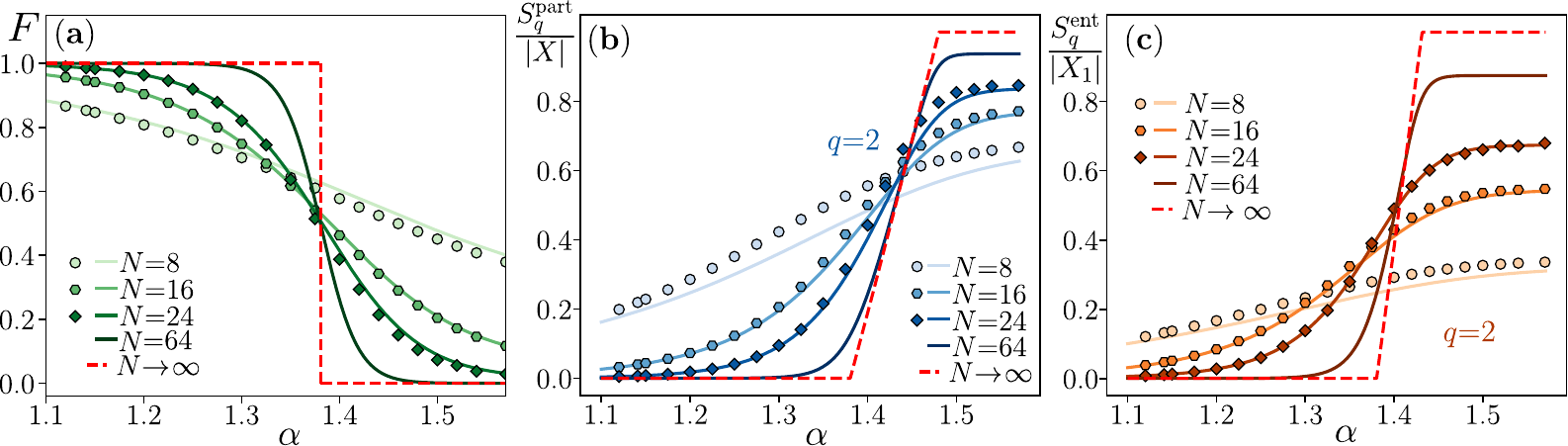}
    \caption{Error-resilience phase transition at code rate $r=k/N=1/4$: (a) the fidelity $F$, (b) the participation entropy $S^{\mathrm{part}}_q$, (c) the entanglement entropy $S^{\mathrm{ent}}_q$ (for subsystem size $|X_1|=N/8$) of the logical state $\ket{\psi^U_X}$. The numerically calculated quenched averages faithfully match the exact analytical expressions at $r=1/4$ as the system size $N$ increases (note that the remaining discrepancies within the available system sizes are larger than the one for $r=1/2$). The EPP to EVP transition is shifted to a larger value of the error strength $\alpha$.
    The error bars are smaller than the marker size.}
    \label{fig:K4}
\end{figure}

Again, we focus on the coherent error case and remark that a similar phenomenology holds also for the depolarizing setup.
In Fig.~\ref{fig:K4}, we report a comparison of numerical results at the code rate $r=1/4$ with the exact analytical expressions for the annealed averages of the fidelity and participation/entanglement entropies. We find qualitative similarities to the $r=1/2$ case in the behavior of all the considered quantities, and we verify the agreement between the analytical expression and numerical results. 

\section{Robustness of the main results to inhomogeneities and further sources of errors }
In the Main Text, we were interested in a constant error rate shared by all the sites. In realistic setups, the quantum noise acts inhomogeneously on different qubits. Furthermore, on noisy quantum devices, even the encoder-decoder circuit can be subject to other sources of noise and errors. 
We touch upon these points in this section. First, we discuss the case of inhomogeneous, site-dependent errors. We show how to extend our analytical methods to case of site-dependent errors and that they allow to pinpoint error-resilience phase transitions. Furthermore, we study the case of noisy encoding and decoding circuits. Our numerical results for that case indicate that the error-resilience transition can be observed for sufficiently weak noise, and that its features disappear is the noise exceeds a threshold value (which depends on the type of the noise).

\subsection{Applications to site-dependent errors}

We now show that the methodologies introduced in this manuscript can be extended to include the inhomogeneous error strength. Specifically, we consider:
\begin{itemize}
    \item the case of coherent errors: we choose $\mathcal{E}_i = e^{-i \alpha_i \sigma^z_i/2} \rho e^{+i \alpha_i \sigma^z_i/2}$, where 
    \begin{equation}
        \alpha_i = \mathcal N(0, W_\alpha)
        \label{eq:alphaDIS}
    \end{equation} are independent Gaussian random variables with zero mean and standard deviation $W_\alpha$ which controls the error strength.
    \item incoherent local depolarizing noise, ${\mathcal{E}_i(\rho) = \sum_{\mu=0}^3 K_{\mu,i}\rho K^{\dag}_{\mu,i}}$, with Kraus operators $K_{0,i} = \sqrt{1-3\lambda_i/4}$, $K_{1,i} = \sqrt{\lambda_i/4} \sigma^x_i$, $K_{2,i} = \sqrt{\lambda_i/4} \sigma^y_i$, $K_{3,i} = \sqrt{\lambda_i/4} \sigma^z_i$, where $\lambda_i$ are taken as independent random variables, each distributed uniformly in the interval 
        \begin{equation}
        \lambda_i \in [0, W_\lambda],
        \label{eq:lambdaDIS}
    \end{equation} 
 where $W_\lambda$ fixes the error strength.
\end{itemize} 
The discussion outlined in the Main Text and in the 
Sec.~"Details on the analytical computations for annealed averages on the full unitary group" of the Supplemental Material extends naturally to the above described case of site-dependent noise. The main difference is how the coefficients $b_\pi(\mathcal{E})$ are computed. 
As already mentioned, the representation of the permutation $\pi$ factorizes on different sites as $T_\tau = t_\pi^{\otimes N}$. It follows that 
\begin{equation}
    b_\pi(\mathcal{E}) = \sum_{\tau} W_{\pi,\tau} \prod_{i=1}^N \mathrm{tr}\left(t_\pi (e^{-i \alpha_i \sigma^z_i/2} \otimes e^{+i \alpha_i \sigma^z_i/2} )^{\otimes q} \right),
\end{equation}
for the coherent errors. Similarly, for the site-dependent local depolarizing noise we find
\begin{equation}
    b_\pi(\mathcal{E}) =   \sum_{\tau,\pi}W_{\tau,\pi}\prod_{i=1}^N \mathrm{tr}\left(t_\pi \left[(1-\lambda_i)\openone_4+ \lambda_i Q_{2} \right]^{\otimes q}\right).
\end{equation}
These formulas imply that the fidelity, at \emph{each fixed} realization of the noise, is given by
\begin{equation}
    \tilde{F} = \frac{\left(2^N-1\right) \left(2^N \prod_{i=1}^N\cos ^2(\alpha_i/2)+1\right)}{2^N \left(2^N-2^k\right) \prod_{i=1}^N\cos ^2(\alpha_i/2)+2^{k+N}-1}\label{eq:falphai}
\end{equation}
for the coherent errors, and, for the site-dependent depolarizing noise, by
\begin{equation}
  \tilde{F} = \frac{\left(2^{N} - 1\right) \left(2^N  \prod_{i=1}^N \left(1 - 3 \lambda_i/4 \right)+1\right)}{ 2^N \prod_{i=1}^N \left(1 - 3 \lambda_i/4 \right) \left(2^{N} - 2^{k}\right)  + 2^{ N + k} - 1  }\label{eq:exlambdF2}.
  \end{equation}
The above formulas are straightforward generalizations of Eq. (5) and (8) of the Main Text. Indeed, in the coherent case, the term $\cos^{2N}(\alpha/2)$ is replaced by $\prod_{i=1}^N\cos ^2(\alpha_i/2)$. For the depolarizing noise $\left(1 - 3 \lambda/4 \right)^{N}$ is replaced by $\prod_{i=1}^N \left(1 - 3 \lambda_i/4 \right)$. 

\begin{figure}[ht]
    \centering
    \includegraphics[width=\columnwidth]{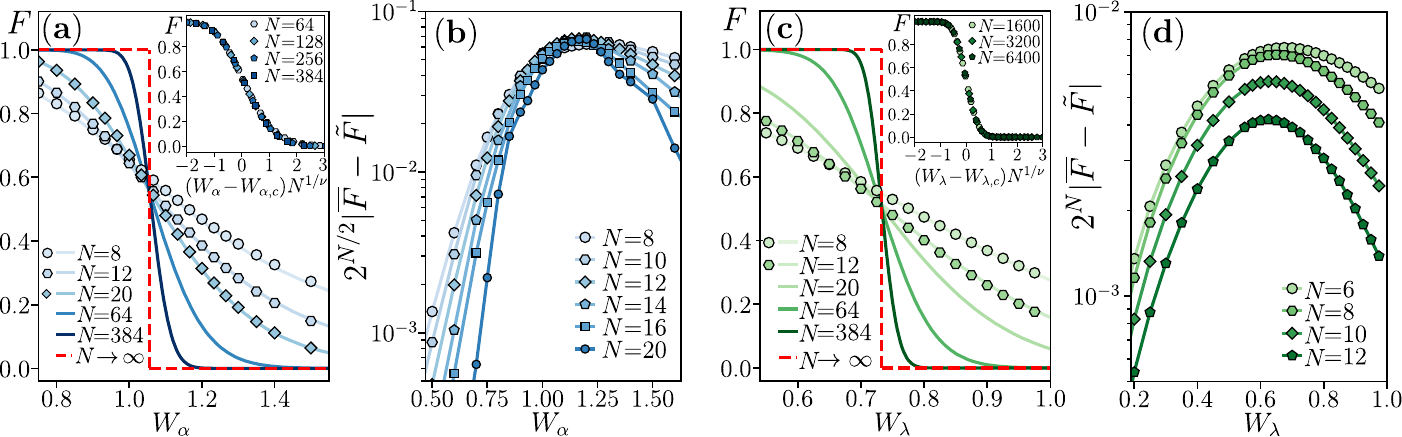}
    \caption{Error resilience phase transition for site-dependent errors. (a) Fidelity of the logical state for coherent rotations by random angles, as a function of $W_\alpha$, see~\eqref{eq:alphaDIS}. Markers show numerically calculated average $\tilde F$, while the lines denote \eqref{eq:falphai} averaged over noise realizations; the inset: collapse of data with $\nu=2$ and $W_{\alpha,c}=1.055(2)$. (b) Rescaled by $2^{N/2}$ modulus of the difference between the quenched and annealed averages of the fidelity $F$ over a sample of over $10^2$ Haar-distributed unitaries $U$ for fixed noise realization $\{ \alpha_j \}_{j=1}^N$, subsequently averaged over more than $10^2$ realizations of the noise of strength $W_\alpha$. (c) Fidelity of the logical state for site-dependent fully depolarizing noise, as a function of $W_\lambda$, see~\eqref{eq:lambdaDIS}. Markers correspond to numerical data, while lines denote \eqref{eq:exlambdF2} averaged over noise realizations; the inset: data collapse with $\nu=2$ and $W_{\lambda,c}=0.7331(2)$. (d) the same as (b) but for the site-dependent depolarizing noise.
    Note the exponential factors multiplying the $|\overline F - \tilde F|$ in panels (b) and (d), highlighting the (super)exponential convergence of $\tilde F$ to $\overline F$.}
    \label{fig:disorder}
\end{figure}

The averaging of results for the site-dependent noise must be done carefully. A "natural" procedure consists of calculation of the logical state for a realization of unitary $U$ and noise $\{ \alpha_j \}_{j=1}^N$ in the coherent case (or $\{ \lambda_j \}_{j=1}^N$ in the incoherent case). We follow this procedure to obtain the numerical results denoted by markers in Fig.~\ref{fig:disorder}~(a),~(c). Considerations analogous to the reasoning presented in Sec.~"Self-averaging in the encoding-decoding circuits" imply that, at each \textit{fixed} realization of the noise (e.g. for the fixed set of $\{\lambda_j\}_{j=1}^N$ in the incoherent case), the fidelity $\tilde{F}$ averaged over the encoding unitaries $U$ is exponentially in $N$ close to the annealed value $\overline{F}$. This property is confirmed numerically in Fig.~\ref{fig:disorder}~(b),~(d), respectively for the coherent and incoherent errors. Consequently, the expressions \eqref{eq:falphai} and \eqref{eq:exlambdF2}, upon averaging over the realizations of the noise lead to results exponentially close in $N$ to the averaged fidelity obtained in the numerics. Even at small system sizes, e.g.~$N\approx 10$, the difference between the annealed and averaged results is remarkably small, see Fig.~\ref{fig:disorder}~(a),~(c).

The results presented in Fig.~\ref{fig:disorder}~(a),~(c) show a transition between EPP and EVP as function of error strength controlled by $W_\alpha$ and $W_\lambda$ respectively for the coherent and incoherent noise. The fidelity approaches a step-function behavior in the thermodynamic limit $N\to \infty$, and the critical exponent is $\nu=2$. The critical exponent differs from the case of homogeneous errors, as the inhomogeneous errors introduce additional disorder to the system which changes the critical behavior at the EPP-EVP transition.


\subsection{The critical exponent $\nu$}

In the Main Text, as well as in the Sec.~"Applications to site-dependent errors" of Supplementary Material, we have presented collapses of the results indicating that the critical exponent for the EPP-EVP transition is $\nu=1$ for the homogeneous error and $\nu=2$ in the site-dependent error case. Below, we present an analytical analysis of the results for the fidelity of the logical state which confirms the values of the exponent $\nu$. The reasoning, outlined for the case of depolarizing noise extends directly to the coherent case.

We are interested in the critical properties of the encoding-decoding circuits, hence, we consider $N \gg 1$ and the code rate $r=k/N$ is fixed as $0<r<1$. For those parameters, keeping only the terms non-vanishing in $N\to \infty$ limit, we find that \eqref{eq:exlambdF2} simplifies to
\begin{equation}
  \tilde{F} =     \frac{\prod_{i=1}^N \left(2 - 3 \lambda_i/2 \right)}{  \prod_{i=1}^N   \left(2 - 3 \lambda_i/2 \right)   + 2^{ r N} }\label{eq:exlambdF2simpl}.
  \end{equation}

Considering first the homogeneous error strength, $\lambda_j \equiv \lambda$, Eq.~\eqref{eq:exlambdF2simpl} can be rewritten as
\begin{equation}
  \tilde{F} =    \left( 1+ 2^{ N (r- \log_2 (2 \left(1 - 3 \lambda/4 \right)) ) } \right)^{-1} \label{eq:exlambdF2simpl2},
  \end{equation}
which readily allows to find that  $\tilde{F}=1/2$ independently of the system size $N$ at the critical error strength $\lambda_c=4(1-2^{r-1})/3$, as provided in the Main Text. Using $\lambda_c$, Eq.~\eqref{eq:exlambdF2simpl2}, becomes
\begin{equation}
  \tilde{F} =    \left( 1+ \left( \frac{1-3\lambda_c/4}{1-3\lambda/4}\right)^N  \right)^{-1} \approx \left( 1+ \exp\left( \frac{N(\lambda-\lambda_c)}{4/3-\lambda_c} \right)  \right)^{-1} \label{eq:exlambdF2simpl3},
  \end{equation}
which demonstrates that the critical exponent is indeed $\nu=1$, as well as determines the shape of the scaling function shown in the inset of Fig.~3(a) in the Main Text.

In the site-dependent case, we introduce a random variable $t$ via $t = \frac{1}{N}\sum_{i=1}^N \log_2 \left(  2 - 3 \lambda_i/2 \right) $. The variable $t$ is an average of $N\gg1$ independent random variables $a_i = \log_2 \left(  2 - 3 \lambda_i/2 \right)$. Hence, by the central limit theorem, the distribution of $t$ is Gaussian, with mean $\mu_t = \overline a_i$ (the overbar denotes an average over the distribution of on-site error strength) and standard deviation $\sigma_t = \overline{ (a_i-\overline a_i)^2}/\sqrt{N}\equiv A/\sqrt{N}$, where $A$ is a constant. The fidelity \eqref{eq:exlambdF2simpl} averaged over realizations of the noise is  
\begin{equation}
 \tilde F = \int_{-\infty}^{\infty} dt
 \frac{\sqrt{N}}{\sqrt{2 \pi A^2}} e^{-\frac{N(t - \mu_t)^2}{2 A^2}}
 \left( 1+ 2^{N(r-t)} \right)^{-1}, \label{eq:int1}
  \end{equation}
where we have used the explicit form of the probability distribution function of variable $t$. The integrand in \eqref{eq:int1} is a product of a Gaussian function of width $\propto 1/\sqrt{N}$ and of the function $\left( 1+ 2^{N(r-t)} \right)^{-1}$ which, as we have seen above, is a step function smeared over an interval of width $\propto 1/N$ and centered at $t=r$. For $N\gg 1$, the latter behavior is much sharper, hence the average fidelity can be approximated as 
\begin{equation}
 \tilde F \approx \int_{-\infty}^{r} dt
 \frac{\sqrt{N}}{\sqrt{2 \pi A^2}} e^{-\frac{N(t - \mu_t)^2}{2 A^2}} 
 = \frac{1}{2} \left( 1+ \mathrm{erf}\left( \sqrt{N} \frac{ r-\mu_t}{\sqrt{2} A}\right) \right), \label{eq:int2}
  \end{equation}
where $\mathrm{erf}$ is the error function. Eq.~\eqref{eq:int2} implies that $ \tilde F = 1/2$ independently of the system size $N$ when $\mu_t = r$. Expressing $\mu_t$ in terms of $W_\lambda$ by calculating the average value $\overline a_i$ using the form of the distribution of $\lambda_i$, \eqref{eq:lambdaDIS}, the condition $\mu_t = r$ becomes a transcendent equation for the critical error strength $W_{\lambda,c}$. The following procedure can be carried out at any $0<r<1$, but we focus at $r=1/2$. Solving the transcendent equation numerically, we find $W_{\lambda,c}=0.7332(1)$, consistently with the result of the preceding section. Then, using a Taylor expansion we readily obtain that $r-\mu_t\approx \alpha_1 (W_\lambda - W_{\lambda,c})$, where $\alpha_1 =0.8864(1)$. This results in 
\begin{equation}
 \tilde F 
 = \frac{1}{2} \left( 1+ \mathrm{erf}\left(  \frac{\alpha_1 }{\sqrt{2} A}  \sqrt{N} ( W_\lambda - W_{\lambda,c} ) \right) \right), \label{eq:int3}
  \end{equation}
which explicitly demonstrates that the critical exponent for the site-dependent error strength is $\nu=2$. Upon plugging in the values of $\alpha_1$ and $A=0.3272(1)$, we find that \eqref{eq:int3} describes accurately the shape of the scaling function presented in the inset of Fig.~\ref{fig:disorder}(c).


\subsection{Noise in encoding-decoding operations}

Now, we turn to the question of whether the error-resilience phase transitions can be observed on noisy quantum devices. So far, the encoding unitary $U$ was either chosen randomly with Haar measure from the unitary group $\mathcal U(2^N)$ (in our analytical calculations), or generated (in the numerical calculations) by a brick-wall quantum circuit built out of $T=2N$ layers of local $2$-qubit gates. In both cases the decoding unitary $U^\dag$ was the inverse of the encoding unitary $U$, and the action of the encoding-decoding circuit was trivial in the absence of errors. This idealized scenario can be, however, realized only approximately on the present-day quantum devices. Indeed, due to the gate errors and noise, the action of the encoding unitary $U$ comprised of local gates will be only approximately undone by $U^{\dag}$. 

To model an experiment with encoding-decoding circuits on a realistic noisy quantum device, we assume that $U$ is a brick-wall circuit composed of the following layers of unitary gates
\begin{equation}
    U_t = \begin{cases}
        \prod_{i=1}^{N/2}U_{2i-1,2i} & \text{if } t \text{ is even},\\
        \prod_{i=1}^{N/2}U_{2i,2i+1} & \text{otherwise},
    \end{cases}
\end{equation}
where $U_{i,j}$ are $2$-qubit gates chosen with the Haar measure the unitary group $\mathcal U(4)$, and  $t=1,\ldots, N$, i.e. the depth of the circuit is $T=N$ (this depth leads to results quantitatively similar to the ones reported above). We posit that each of the unitary layers $U_t$ is followed by a layer of single qubits depolarization channels, ${\mathcal{E}_i(\rho) = \sum_{\mu=0}^3 K_{\mu,i}\rho K^{\dag}_{\mu,i}}$ with Kraus operators $K_{\mu,i}$ ($\mu=0,1,2,3$) given in the Main Text, and the local error strengths $\lambda_i$ are chosen to be $\lambda_i=\varepsilon$ with probability $1/4$ and  $\lambda_i=0$ otherwise. This results in a channel $\rho \to \sum_{\vec \mu} K_{\epsilon,\vec{\mu}} \rho K_{\epsilon,\vec{\mu}}^\dag$ specified by $4^N$ Kraus operators $K_{\epsilon,\vec{\mu}}= \prod_{j=1}^N K_{\mu_j,j}$, where $\vec{\mu} = ( \mu_1, \ldots \mu_N )^T$. 
After $T=N$ unitary layers $U_t$ followed by the depolarizing errors, there is a layer of local coherent errors $N_{\alpha} = \prod_{j=1}^{N}e^{-i \alpha \sigma_j^z/2}$ of strength $\alpha$ which allows us to tune the system across the EPP-EVP transition, and finally a brick-wall circuit that models the decoding unitary $U^{\dag}$.
Taking $ \rho_0=(\ket{ 0 }\bra{0 })^{\otimes N}$ as the initial state, the final state reads 
\begin{equation}
\rho_\varepsilon = 
\sum_{\vec{\mu}_1, \ldots, \vec{\mu}_{2T}}K_{\epsilon,\vec{\mu}_{2T} } U^\dag_T\ldots K_{\epsilon,\vec{\mu}_{T+1}} U^\dag_1
    N_{\alpha} K_{\epsilon,\vec{\mu}_T} U_T\ldots K_{\epsilon,\vec{\mu}_1} U_1  \rho_0 U^\dag_1 K^{\dag}_{\epsilon,\vec{\mu}_1} \ldots U^\dag_T K^{\dag}_{\epsilon,\vec{\mu}_T} N^{\dag}_{\alpha} U_1 K^{\dag}_{\epsilon,\vec{\mu}_{T+1}} \ldots U_T K^{\dag}_{\epsilon,\vec{\mu}_{2T} }.
    \label{eq:errU}
\end{equation}
Subsequently, the final decoded state is obtained by projecting out the ancilla qubits: ${\rho_{\varepsilon,X} = \langle 0_{\bar{X}}|\rho_\varepsilon|0_{\bar{X}}\rangle/\mathrm{tr}(\langle 0_{\bar{X}}|\rho_\varepsilon|0_{\bar{X}}\rangle)}$. 

\begin{figure}[ht]
    \centering
    \includegraphics[width=\columnwidth]{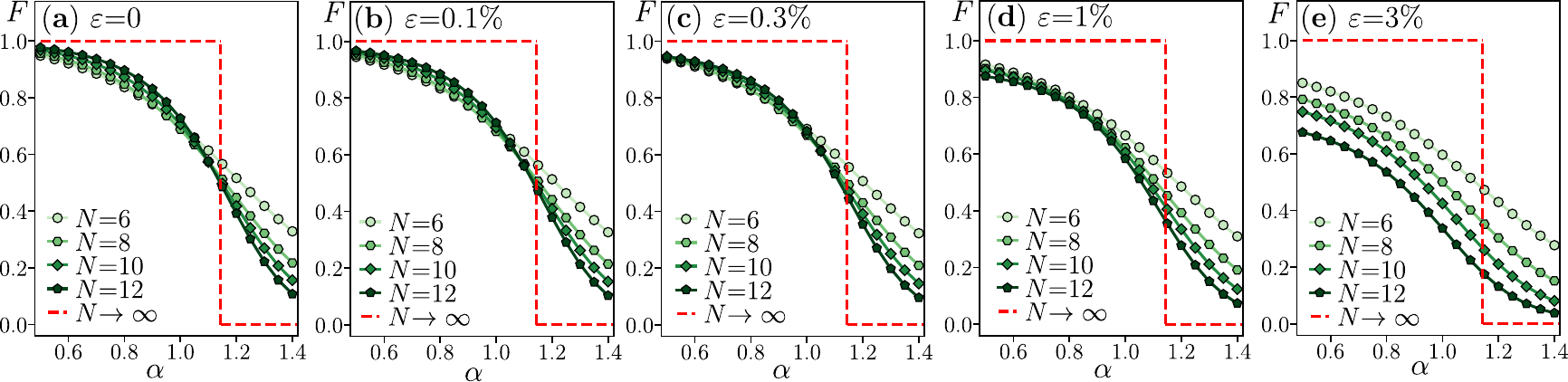}
    \caption{Error resilience phase transition in presence of depolarizing noise of strength $\varepsilon$ in the encoding and decoding unitaries modeled by \eqref{eq:errU}. The fidelity $F$ is shown as a function of the coherent error strength $\alpha$ for various system sizes $N\in[6,12]$. The transition is observable for $\varepsilon > 0$, however, its features are gradually smeared when the noise strength is increased. The red dashed lines shown the analytical prediction for the fidelity in noiseless system ($\varepsilon=0$) in thermodynamic limit $N \to \infty$. }
    \label{fig:errU}
\end{figure}

For $\varepsilon = 0$, this setup reproduces the ideal encoding-decoding circuit with coherent errors discussed Main Text, see Fig.~2. Consequently, at $\varepsilon=0$, we observe the transition between EPP and EVP, see Fig.~\ref{fig:errU}(a). This behavior of the fidelity $F= \langle 0 |^{\otimes k}\rho_{\varepsilon,X}|0\rangle^{\otimes k}$ persists also when $\varepsilon >0 $. Indeed, for $\varepsilon = 0.1 \%$, the behavior of $F$ as a function of $\alpha$ is practically unaltered by the presence of the noise, as shown in Fig.~\ref{fig:errU}(b). We clearly observe an error-protecting regime in which $F$ increases towards unity with increasing $N$, as well as an error-vulnerable regime, in which the fidelity $F$ decreases towards $0$. This indicates that the error-resilience phase transitions can be observed on present-day noisy quantum devices, provided that the noise and gate errors are sufficiently small. In particular, for such a weak noise even the critical content of the EPP-EVP can be probed on a noisy device. Importantly, however, due to the progressive accumulation of the errors with increasing spatio-temporal volume of the circuit, the prospects of experimental observation of the transition diminish with increasing $N$. We expect that the fidelity $F$ eventually decays to $0$ at sufficiently large $N$, even at $ \varepsilon = 0.1 \%$. As Fig.~\ref{fig:errU}(c) demonstrates, the error-protecting and vulnerable regimes can still be clearly discriminated at the investigated system sizes at $\varepsilon=0.3 \%$. Further increase of the noise strength leads to deterioration of the fidelity $F$. At $\varepsilon=1\%$, we observe that $F$ is monotonically decreasing with $N$ in the whole studied interval $0.6 \leq \alpha \leq 1.4$, cf. Fig.~\ref{fig:errU}(d). Nevertheless, even at $\varepsilon=1\%$, we observe a sharpening of the $F(\alpha)$ curve with increasing $N$ which is still an observable premise of the EPP-EVP transition. Finally, the features of the EPP-EVP transition become completely obscure once the noise gets sufficiently strong and fidelity deteriorates at any $\alpha$, see Fig.~\ref{fig:errU}(e).

In conclusion, we have analyzed a simple model \eqref{eq:errU} of encoding-decoding circuits on a noisy quantum device. Our numerical analysis indicates that EPP-EVP transition, including its critical features, can be observed on a noisy device provided that the noise is sufficiently weak. We note that we have not performed here any optimization of the parameters of the circuit. For instance, by decreasing the circuit depth, we could diminish the effects of gate errors and noise. Moreover, for a specific noisy quantum device, one could employ a more realistic model of noise and use error mitigation techniques to further improve the observability of the error resilience phase transitions.

\end{document}